\documentclass{article}
\usepackage{graphicx}
\usepackage{amsmath,amssymb}
\usepackage{bm}
\usepackage{bbm}
\usepackage{natbib}
\usepackage{booktabs}

\usepackage{cleveref}[noabbrev]
\usepackage{geometry}
\usepackage{ulem}
\usepackage{lineno}
\usepackage{tikz}
\usetikzlibrary{shapes, arrows}
\newcommand{\R}{\mathbb{R}}

\newcommand{\bx}{\bm{x}}

\newcommand{\E}{\text{E}}
\newcommand{\cov}{\text{Cov}}
\newcommand{\var}{\text{Var}}

\newcommand{\diag}{\text{diag}}

\let\oldequation\equation
\let\oldendequation\endequation

\renewenvironment{equation}
  {\linenomathNonumbers\oldequation}
  {\oldendequation\endlinenomath}
\renewenvironment{equation*}
    {\linenomathNonumbers\oldequation}
    {\oldendequation\endlinenomath}
\usepackage{verbatim}
\usepackage{xcolor}
\newcommand{\dah}[1]{\textcolor{blue}{#1}}

\begin{document}
%
\title{Multilevel Emulation for Stochastic Computer Models with Application to Large Offshore Wind Farms}
\author{Jack C. Kennedy\footnote{Corresponding author: \texttt{j.c.kennedy1@ncl.ac.uk}}, Daniel A. Henderson and Kevin J. Wilson
\\ \small School of Mathematics, Statistics and Physics, Newcastle University, UK \normalsize }
%
\maketitle
\begin{abstract}
Renewable energy projects, such as large offshore wind farms, are critical to achieving low-emission targets set by governments. Stochastic computer models allow us to explore future scenarios to aid decision making whilst considering the most relevant uncertainties. Complex stochastic computer models can be prohibitively slow and thus an emulator may be constructed and deployed to allow for efficient computation. We present a novel heteroscedastic Gaussian Process emulator which exploits cheap approximations to a stochastic offshore wind farm simulator. We also conduct a probabilistic sensitivity analysis to understand the influence of key parameters in the wind farm model which will help us to plan a probability elicitation in the future.
\end{abstract}
\section{Introduction}
Offshore wind farms are becoming an increasingly attractive approach to the generation of clean, renewable energy \citep{Hobley2019}. To exploit the abundance of offshore wind, wind farms are utilising increasing numbers of turbines. For example, the world's largest offshore winds farms (measured by number of turbines) are the London Array with $175$ turbines and the Hornsea $1$ which has $174$ \citep{Paterson2018}. Also, new technologies and placement of the turbines further away from the coast in new, harsh, deep-water environments induces a large number of uncertainties about, for example, the lifetimes of critical components. This ultimately impacts energy generation and profits. Uncertainty needs to be investigated prior to investing time and money into the development of highly ambitious renewable energy projects. Stochastic computer modelling is a cost-effective approach to exploring future scenarios, but is not without its own challenges.

In this paper, we focus on the Athena simulator \citep{Zit13, Zit16}, a stochastic point process model of an offshore wind farm. The main purpose of the Athena simulator is decision support under uncertainty. The uncertainties considered in Athena are the epistemic uncertainty about simulator parameters and the aleatory uncertainty about the natural world, for instance, the weather.

For example, an engineer designing the wind farm may be able to choose between a tried and tested component or a novel design. This could be a choice between one of two gearboxes. The engineer would formulate uncertainty distributions over parameters governing gearbox performance and then propagate this uncertainty through Athena to understand how the uncertainty in component performance impacts wind farm performance. When eliciting parameters for a complex computer model, such as the Athena simulator, it is not clear which parameters are most important until a sensitivity analysis has been conducted. Probabilistic sensitivity analysis (PSA) allows us to quantify the proportion of output uncertainty (measured by variance) induced by any input. The most important inputs are those contributing the most to output uncertainty \citep{Oakley04}.

A bottleneck we encounter is that the Athena simulator is computationally expensive, thus PSA becomes infeasible. The stochastic nature of Athena makes these computations even more cumbersome. An effective approach in such scenarios is to build a fast statistical surrogate model --- an \textit{emulator} --- to replace the simulator \citep{Sacks89,Gramacy2020surrogates}, thus making PSA feasible.

There are a variety of approaches to emulation of stochastic computer models; see \citet{Baker2020} for a recent overview. A desirable feature of these emulators is that they give a mean response and a quantification of both types of uncertainty in the simulators; the epistemic uncertainty quantifies our uncertainty about mean simulator output, and the aleatory uncertainty quantifies the simulator's level of noise at any tried or untried input.
Many Gaussian process (GP) based emulation approaches for stochastic problems rely on large levels of replication, which is appropriate when a sufficiently large computing budget is available for training data; see \citet{Henderson09,Akenman2010,Plumlee2014} or \citet{Andrianakis2017}. Athena can take up a prohibitive amount of time for a single accurate run, thus such levels of replication would make emulation of the Athena model infeasible.

An approach which need not require replication, but still allows for it, is the heteroscedastic GP (HetGP) \citep{Goldberg1998, hetGP}. The allure of HetGP is the promise of a full surrogate; joint prediction of the mean response and the noise level at any input combination. This is possible via a latent variable formulation which jointly models the simulator mean and the log noise (to ensure positivity) as GPs. As \citet{Gramacy2020surrogates} notes, this coupled GP approach provides smooth estimates of the noise at both within sample and out of sample simulator inputs. This very flexible approach to emulation is incredibly data-hungry. For example, \citet{Binois2018} use $500$ design points to compare emulators for a one dimensional stochastic simulator.

In this paper, we exploit the flexibility of HetGP for emulating the stochastic Athena simulator. We also seek to circumvent the data-hungry nature of HetGP by exploiting the simplicity with which we can change model features within the Athena simulator to give us cheap approximations. Since these approximations are fast, it is easier to construct good emulators. If we can build a good emulator for the cheap simulator, and accurately describe its mean, we can utilise this information to build better emulators for more expensive stochastic computer models.

Exploiting cheap approximations to an expensive simulator has been tackled in the deterministic framework by \citet{Kennedy2000}. The most popular format is their autoregressive structure for functions \citep{Forrester2007, Singh2017, Harvey2018}. The autoregressive structure builds a well informed emulator for the cheap simulator and uses this as a ``starting point'' for the expensive simulator. The main aim of multilevel emulation is an improved emulation of the simulator at a fixed training budget. We extend this to the more complex case of stochastic computer experiments to enhance the emulation of the Athena simulator.

The remainder of the article is structured as follows. Section~\ref{sec:athena} provides some relevant background information on the Athena simulator and Section~\ref{sec:hetgp} provides a brief overview of emulation via heteroscedastic Gaussian processes. In Section~\ref{sec:SML} we present mathematical details of stochastic multilevel emulation, which is a key contribution of this article. Section~\ref{sec:SML-athena} constructs and compares emulators for Athena. Probabilistic sensitivity analysis is performed in Section~\ref{sec:PSA}  and Section~\ref{sec:conc} contains concluding remarks.
\section{Athena: a stochastic model of a wind farm}
\label{sec:athena}

The Athena simulator is a point process model of a wind farm which simulates events at discrete times over a time period $[0, T_{max}]$. Events are a component in the wind farm being damaged or repaired. Events can also be the triggering of farm-wide maintenance or the deployment of a boat to perform a repair. To simulate events, the simulator starts from time $t=0$ and calculates the hazard function of each event at each time point over the period of interest; this is a function of time and the state of the wind farm. From this the ``total'' hazard (at each time point), known as the Force of Mortality (FOM), is calculated which then implies the next event time. If we are at $T_{p-1}$ then the time to the next event, $T_p$ is found as $R^{*}(T_p) = R^{*}(T_{p-1}) + E$ where $E \sim Exp(1)$ and $R^{*}$ is the cumulative intensity function of the wind farm. This time $T_p$ is the solution to an integral which depends on the wind farm's current state, which is part of a stochastic process.
The integral is
\begin{equation}
R^{*}(\tau_n) - R^{*}(\tau_{n-1}) = \int_{\tau_{n-1}}^{\tau_n} r^{*}(u) \textrm{d} u  \label{Eq:intensity}
\end{equation}
where $r^{*}$ is the wind farm's intensity function. If $\tau_{n-1}=T_{p-1}$ is the time of the last event, then $R^{*}(\tau_{n-1})$ is known. The goal is to find $R^{*}(\tau_n)$ by increasing $\tau_{n}$ sequentially to $T_{p-1} + \Delta t$, $T_{p-1} + 2\Delta t$, \ldots until $R^{*}(\tau_n) - R^{*}(\tau_{n-1}) > E$. The first value of $\tau_{n}$ satisfying the inequality is taken as $T_p$. Athena then decides which subassembly caused the event. If $y_{j,k}(T_p)$ is an indicator taking the value $1$ when subassembly $(j,k)$ has failed and zero otherwise, $\lambda_{j,k}(T_p)$ is the failure intensity of the subassembly and $\mu_{j,k}(T_p)$ is its restoration intensity, then the probability that subassembly $(j,k)$ caused the event at time $T_p$ is
\begin{equation}
  p_{j,k}(T_p) = \frac{y_{j,k}(T_p)\lambda_{j,k}(T_p) + (1-y_{j,k}(T_p))\mu_{j,k}(T_p)}{\sum_{j,k} \left\{y_{j,k}(T_p)\lambda_{j,k}(T_p) + (1-y_{j,k}(T_p))\mu_{j,k}(T_p)\right\}}.
\end{equation}
These probabilities form a partition of $[0,1]$, so drawing a $U(0,1)$ random variable allows us to simulate which event occurred. This is repeated until we reach the end of the pre-specified simulation period $T_{max}$.

The Athena simulator models the states of the ``sub-assemblies'' of each turbine in the wind farm. In particular, it models the turbines as being constructed of $8$ main subassemblies and a $9$th `catch all' subassembly which collectively models the behaviour of several unimportant components which together have a non-negligible effect. The subassemblies are the gearbox, generator, frequency converter, transformer, main shaft bearing, the blades, tower,
foundations and the catch all.
The time to failure, $T_{j,k}$, of subassembly $j$ in turbine $k$ is modelled by a non-stationary Weibull distribution: $T_{j,k} \sim Weibull( \alpha_{j,k}(t), \kappa_{j,k}(t) )$. The hazard for a subassembly follows a `bathtub' hazard function which controls $\alpha_{j,k}(t)$ and $\kappa_{j,k}(t)$ . A bathtub hazard function corresponds to three main stages of component life (i) infant mortality in which a larger than expected number of components fail due to manufacturing faults (decreasing hazard); (ii) useful life in which a component works as expected (constant hazard); (iii) degradation in which a component is beyond its useful life (increasing hazard). Athena incorporates many extra details into the hazard function. Performing maintenance tasks extends the expected life of a subassembly, whereas operator misuse decreases lifetimes. A concept known as `virtual life' allows us to replace a completely broken component with a new one whilst keeping the indices $(j,k)$ unchanged. A driver of subassembly lifetime is the onset of ageing, that is, the start of phase (iii) of the hazard function.

\begin{figure}[ht]
	\centering
	\includegraphics[width = 0.75\textwidth]{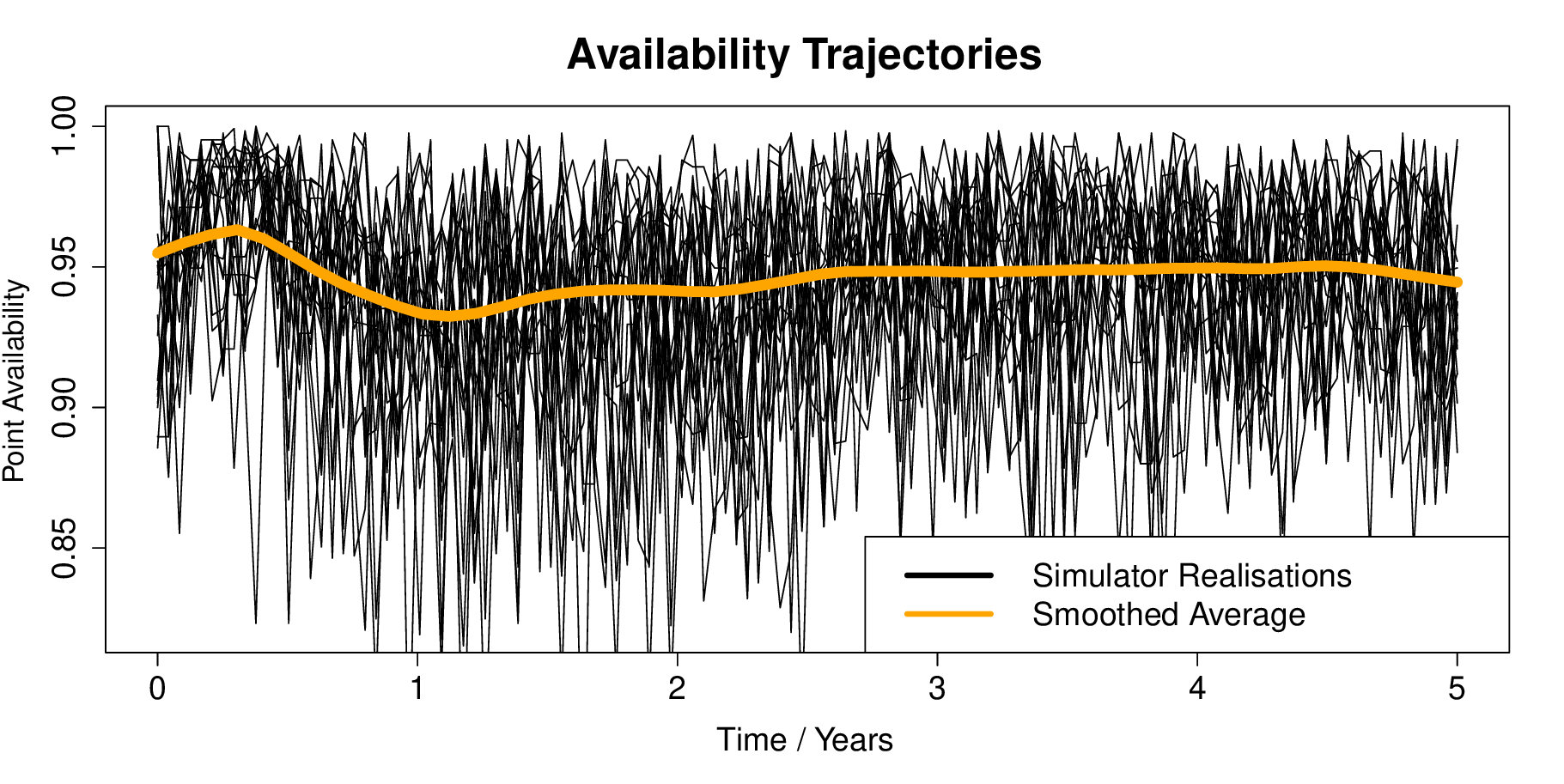}
	\caption{A collection of $10$ availability trajectories (black lines) over the first $5$ years of a wind farm's operational life for a fixed set of parameter values. The orange line represents a smoothed average of the trajectories.}
	\label{Fig:availability-trajectories}
\end{figure}

 In practice, the values of many model parameters are unknown thus uncertainty distributions are to be elicited from experts and propagated through Athena to understand how input uncertainty induces uncertainty in key metrics.
A key model output is a time series which tracks the ``availability'' of a wind farm over time (see \Cref{Fig:availability-trajectories}). Availability is a measure of reliability (performance) of offshore wind farms; the availability at time $t$ is the energy output of the wind farm as a proportion of the maximum possible energy output at time $t$. We compress the time series into a single value --- the mean availability. Offshore wind farms reach an availability of around $93\%$ for near shore turbines, but this is reduced for turbines further away from the coast since reaching the turbines for repair is much more difficult \citep{Carroll2016}. Availability is related to a wind farm's uptime and hence its profitability.
 In the first $5$ years of operation, \textit{excessive failure} is frequently observed. That is, the wind farm typically under-performs due to higher than expected numbers of component failures; tackling this issue is vital to the feasibility of offshore wind.

\section{Heteroscedastic Gaussian Processes (HetGP)}
\label{sec:hetgp}
\begin{figure}[ht]

  \label{Fig:log-vars}
  \centering
  \includegraphics[width=0.75\textwidth]{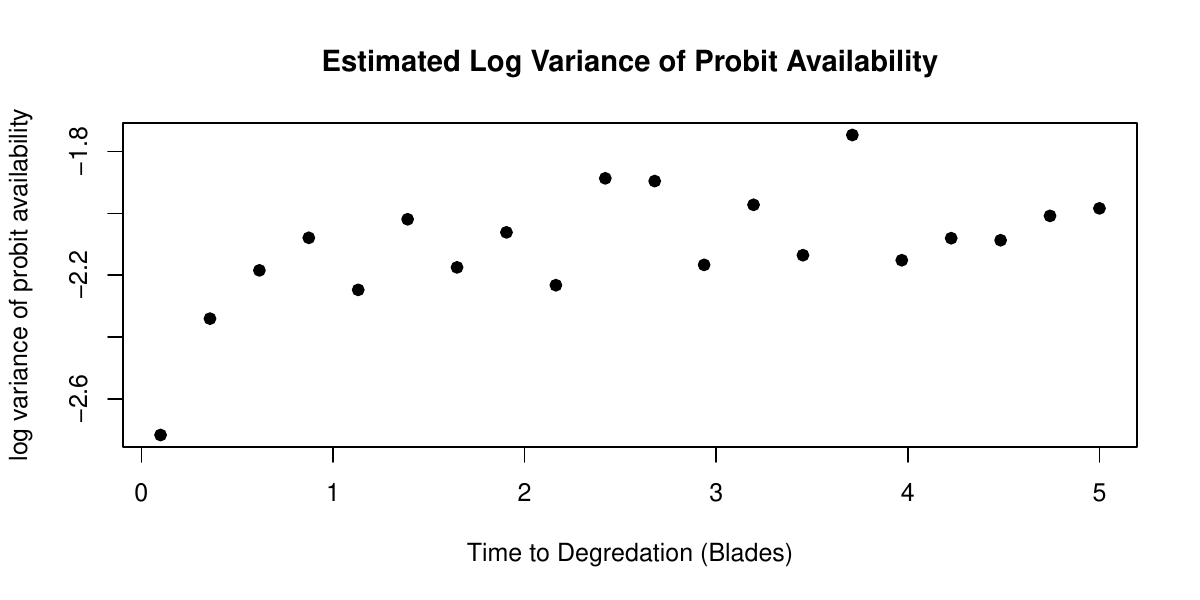}
    \caption{Log sample variances of the probit mean availability. }
\end{figure}
One challenging aspect of the Athena simulator is heteroscedasticity. \Cref{Fig:log-vars} shows (log) sample variances of probit mean availability plotted against the time to degradation of the blades. The probit transformation is chosen because availability is constrained to the unit interval but GPs are defined on the real line. Even after transformation, heteroscedasticity is present and thus should be modelled. We therefore outline HetGP \citep{Binois2018} to later draw parallels with Stochastic Multilevel (SML) emulation in Section~\ref{sec:SML}.

Suppose we have a complex stochastic simulator, $\eta(\cdot)$. We can model this as a HetGP,
\begin{align*}
\eta(\cdot) | \lambda^2(\cdot) &\sim  \mathcal{GP} \{ m(\cdot), C(\cdot, \cdot) + \lambda^2(\cdot) \}\nonumber \\
\log \lambda^2 (\cdot) &\sim  \mathcal{GP} \{ m_V (\cdot), C_V (\cdot, \cdot) + \lambda_{V}^2 ) \}.\nonumber
\end{align*}
\noindent Here, $m(\cdot)$ and $m_V(\cdot)$ are prior mean functions for the simulator's mean and log variance, respectively. The mean functions are expressed in a hierarchical form; $m(\bx) = h(\bx)^T \bm{\beta}$, where $\bx$ is the $K$ dimensional simulator input. The vector $h(\bx)$ is a collection of simple, deterministic basis functions \citep{Fricker2011, Becker2012} and $\bm{\beta}$ are unknown coefficients to be inferred. The mean function on the log variance is expressed similarly; $m_V(\bx) = h_V(\bx)^T \bm{\beta}_V$. Here, $\lambda^2(\cdot)$ is the noise of the expensive simulator; $\log \lambda^2(\cdot)$ is modelled by a GP which itself has noise $\lambda^2_V$. Since the noise (and therefore covariance function) depends explicitly on $\bx$, HetGP is a type of non-stationary GP.
 A common choice of covariance function for computer experiments is the squared exponential covariance function, as this imposes the belief that the moments of the simulator output are smooth functions of the simulator inputs \citep{Santner2003}. A squared exponential covariance function, for a simulator with $K$ inputs, is of the form $C(\bx, \bx') = \sigma^2 \exp \left\{  -(\bx - \bx')^{T}D^{-1}(\bx - \bx') \right\}$, where $\sigma^2$ is a scale parameter and $D = \text{diag}(\theta_1 ^2, \ldots, \theta_K ^2)$ is a diagonal matrix of correlation lengthscales. The same form is given to $C_V$, but the parameters ($\sigma_V^2$, $\theta_{k, V}$) can take different values.
The simulator is run $n$ times to obtain training data $\mathcal{D} = \{y_i, \bx_i : i = 1, \ldots, n\} $, where $y_i$ are runs of $\eta(\bx_i)$. The hyperparameters, \newline $\Theta = \{\theta_1, \ldots, \theta_K, \theta_{1,V}, \ldots, \theta_{K,V}, \bm{\beta}, \bm{\beta}_V, \sigma, \sigma_V, \lambda_{V} \}$, are inferred and the log variance, $\log \lambda^2(X) = \left( \log \lambda^2(\bx_1), \ldots, \log \lambda^2(\bx_n) \right)$, at the design points, $X = \{\bx_1, \ldots, \bx_n \}$, can be estimated.  We take an empirical Bayes (EB) approach to estimation as a compromise between (i) computational cost and (ii) comprehensive uncertainty quantification. A fully Bayesian approach, via MCMC, allows us to quantify and propagate uncertainty about all unknowns but is highly computationally expensive \citep{Kersting2007}. A point estimate, such as a \textit{maximum a posteriori} (MAP) estimate, is faster to compute but we found has poor uncertainty quantification when a non-constant mean function is used. The EB approach offers analytic uncertainty quantification in the $\beta$ parameters. Further, we found the EB estimate of the unknown parameters to be about $60$ times faster to fit than a MAP estimate due to the reduced parameter space.

 We assign priors $\bm{\beta} \sim N(\bm{b}, B)$ and $\bm{\beta}_{V} \sim N(\bm{b}_{V}, B_{V})$, marginalise out the $\beta$ coefficients and obtain a MAP estimate of the GP covariance structure. After integrating out $\bm{\beta}_V$ we can write the joint density of $\log \lambda^2(X)$ and $\log \lambda^2(X^\star)$, where $X$ and $X^{\star}$ are collections of simulator inputs, as
\begin{equation}
  \begin{pmatrix}
    \log \lambda^2(X) \\
    \log \lambda^2(X^\star)
  \end{pmatrix} \mid \Theta_{-\beta}\sim \mathcal{N}\left\{
    \begin{pmatrix}
      H_V \bm{b}_V \\
      H^{\star}_V \bm{b}_V
    \end{pmatrix},
    \begin{pmatrix}
        K_V (X, X) + \lambda^2_V I  & K_V(X, X^{\star})\\
        K_V(X^{\star},X) &   K_V (X^{\star}, X^{\star}) + \lambda^2_V I
    \end{pmatrix}
   \right\}
\end{equation}
where $\Theta_{-\beta}$ denotes the vector $\Theta$ with $\bm{\beta}$ and $\bm{\beta}_V$ removed. We have also introduced $K_V(\bx, \bx') = C_V(\bx, \bx') + h_V(\bx)^T B_V h_V(\bx')$; the covariance between two log variances after integrating out $\bm{\beta}$ and $\bm{\beta}_V$. Finally, $H_V$ is the design matrix for the log variance of simulator inputs and $H^{\star}_V$ is the design matrix for the log variance at some untried inputs $X^{\star}$.

Conditional on $\mathcal{D}$, $\Theta_{-\beta}$, and $\log \lambda^2(X)$, the posterior predictive distribution of $\log \lambda^2 (X^\star)$ is
\begin{equation}
\log \lambda^2(X^\star) \mid \log \lambda^2(X), \mathcal{D}, \Theta_{-\beta} \sim \mathcal{N} \left\{ m_V^\star(\bx^\star), K_V^\star(\bx^\star, \bx^\star) + \lambda_{V}^2 \right\}, \nonumber
\end{equation}
where the posterior moments are found via the conditional normal equations,
\begin{align}
m^\star_V(X^\star) & = H^{\star}_V \bm{b}_V + K_V (X^\star, X)\left[K_V(X, X) + \lambda_{V}^2 I_n \right]^{-1}\left(\log \lambda^2(X) -  H_V \bm{b}_V \right) \nonumber\\
K_V^\star(X^\star, X^\star) & = K_V(X^\star, X^\star) - K_V(X^\star, X) \left[K_V(X, X) + \lambda_{V}^2 I_n \right]^{-1}K_V(X, X^\star)\nonumber
\end{align}
and $I_n$ is the $n \times n$ identity matrix.
Now the joint density for the observed simulator outputs $\eta(X)$ and the output $\eta(X^{\star})$ at new inputs $X^{\star}$ is
\begin{equation}
  \begin{pmatrix}
    \eta(X) \\ \eta(X^\star)
  \end{pmatrix}\mid \begin{pmatrix}\Theta_{-\beta} \\ \lambda^2(X)  \\ \lambda^2(X^\star)\end{pmatrix} \sim \mathcal{N}\left\{ \begin{pmatrix}
  H \bm{b} \\ H^\star \bm{b}
\end{pmatrix}, \begin{pmatrix} K(X, X) + \lambda^2(X)I & K(X, X^\star)\\
K(X^\star, X) & K(X^\star, X^\star) + \lambda^2(X^\star)I
\end{pmatrix} \right\}.
\end{equation}
Here we estimate $\lambda^{\star 2} (X^{\star} )$ with $\exp\{ m_V ^\star (X^{\star} ) \}$. We determine $m^\star(X^{\star} )$ and $K^{\star}(X^{\star} , X^{\star} )$ by the conditional normal equations,
\begin{align*}
m^{\star}(X^{\star}) &= m(X^{\star} ) + K( X^{\star} , X) \{ K(X,X) + \lambda^{\star 2} (X) I \}^{-1} ( \bm{y} - m(X) ) \\
K^{\star}(X^{\star} , X^{\star} ) &= K( X^{\star} , X^{\star} ) -  K(X^{\star} , X) \{ K(X,X) + \lambda^{\star 2} (X) I \}^{-1} K(X, X^{\star} ),
\end{align*}
where $\bm{y} = (y_1, \ldots, y_n)^T$.

\begin{figure}[ht]
	\centering
	\includegraphics[width=0.75\textwidth]{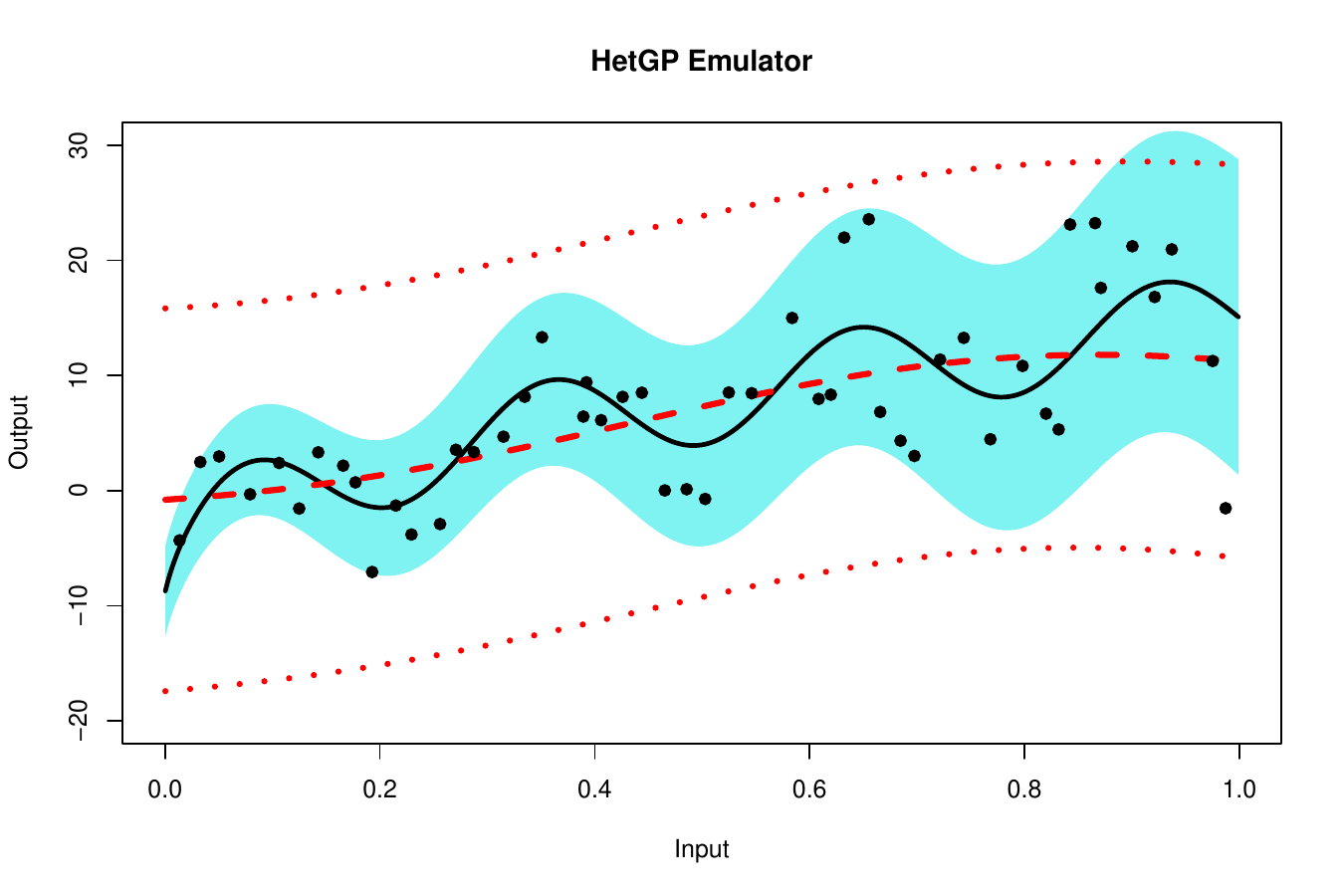}
	\caption{A HetGP emulator for $\eta(\cdot)$. Black points are the outputs from $50$ runs of the simulator. Black line represents the true simulator mean and the blue band represents the mean $\pm 1.96$ `true' standard deviations. Red dashed line represents emulator mean with red dotted lines being the emulator mean $\pm 1.96$ emulator standard deviations.}
	\label{Fig:toy-HetGP}
\end{figure}

In \Cref{Fig:toy-HetGP} we see an example HetGP emulator for the stochastic simulator $\eta(x) = 4\sin(7 \pi x) + 5(2x + 1) + 3 \log (x + 0.01) + (5x + 2)\varepsilon$ where $\varepsilon \sim \mathcal{N}(0,1)$ and $x \in [0,1]$.
Observing the fit in \Cref{Fig:toy-HetGP}, the fitted emulator mean (dashed red line) does not match up well with the simulator. The emulator predicts an approximately linear response whereas the simulator is clearly sinusoidal in nature. The emulator is interpreting the systematic sinusoidal variation as noise, rather than signal. Ultimately, this is because emulating a stochastic computer model requires much more information than the standard deterministic problem. However, when provided with adequate amounts of data, HetGP can produce excellent surrogates for complex stochastic computer models \citep{Binois2018}.

\section{Stochastic multilevel emulation}
\label{sec:SML}

\subsection{Motivation and intuition}

In this section, we outline our proposed approach to stochastic multilevel (SML) emulation of stochastic simulators. This naturally extends deterministic emulation techniques and exploits the cheap approximations that are readily available from the Athena simulator. This approach is quite general and will apply to many stochastic simulators when cheap approximations are available. Many stochastic computer simulators have a complexity parameter, such as the length of a time step, or granularity of a grid over space, which exchanges simulation accuracy for computational cost; examples include \cite{Kennedy2000} and \citet{Le2014}. In our wind farm setting this will be the time step, $\Delta t$, in a numerical integration within each simulation run.

 The number of event times is affected by $\Delta t$, which generates the random time between events. Accurate runs ($\Delta t = 0.001$) of the Athena simulator take just over $3$ minutes for a wind farm with $200$ turbines on a desktop PC with $8 \times 3.20 \,\text{GHz}$ processors and $16\,\text{GB}$ RAM. On the same machine, cheap runs ($\Delta t = 0.1$) take just under $3$ seconds. A single expensive run is computationally equivalent to $60$ cheap runs.The accuracy required comes at a computational cost which severely hinders the size of our computer experiment, limiting the quality of the fitted emulator. We aim to exploit these computational properties in jointly modelling the ``cheap'' simulator and ``expensive'' simulator. The outputs from cheap and expensive versions of stochastic simulators will be related. Runs from both versions are combined to build an overall better emulator.

\begin{figure}[ht]
  \centering
  \includegraphics[width=0.75\textwidth]{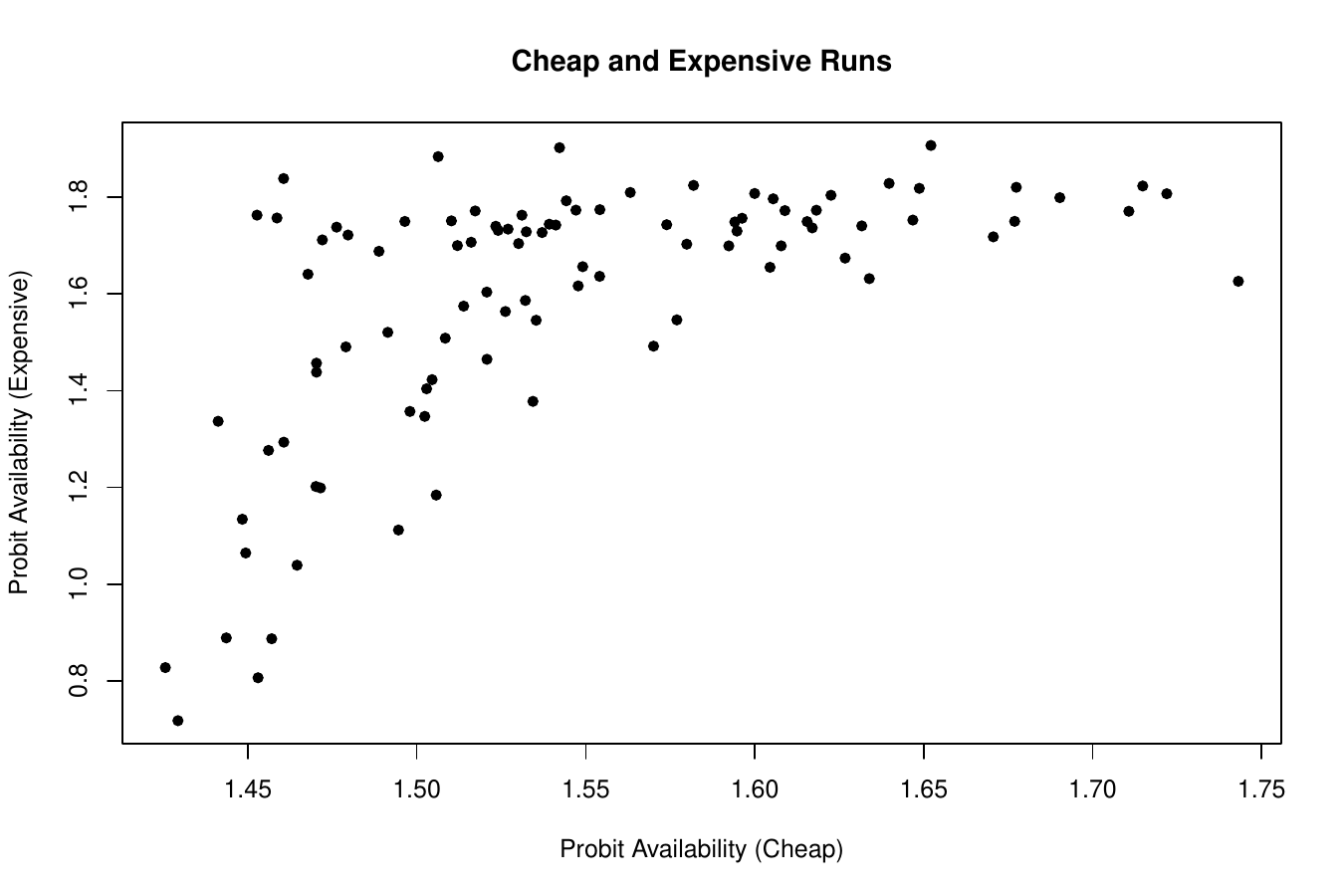}
\caption{Mean probit availability under the cheap version plotted against the expensive version. Each point is computed via $10$ replications. We see an approximately linear relationship between the two levels of code but note that the range of the axes is quite different: $0.7<\text{Expensive}<2$ but $1.4<\text{Cheap}<1.8$. }
  \label{Fig:cheapandexp}
\end{figure}

 The two levels of the Athena simulator are approximately linearly related; see \Cref{Fig:cheapandexp}. The relationship is not exact, partially due to the stochasticity of the two levels. The relationship flattens off when the probit cheap code exhibits values above about $1.6$.
We will focus on a two level set up;  $\eta^C(\cdot)$ is the cheap simulator and $\eta^E(\cdot)$ is its expensive counterpart. In the motivating example of the Athena simulator $\eta^C(\cdot)$ is a version of the model with a time step of $\Delta t = 0.1$ years (simulating time steps of just over a month). However, we want to infer $\eta^E (\cdot)$, which is a version with time step $\Delta t = 0.001$ years (simulating time steps of approximately $9$ hours).
\subsection{Proposed emulation strategy}
We allow for $\eta^E(\cdot)$ to be heteroscedastic but if we believe it is homoscedastic we can replace the non-constant variance with a constant term. Our object of inference is (the distribution of) $\eta^E(\bx)$, for any $\bx$.

Suppose that the cheap simulator, $\eta^C(\cdot)$ can be modelled by a homoscedastic (constant noise) GP with mean function $m_C(\cdot)$, covariance function $C_C(\cdot, \cdot)$ and constant variance $\lambda^2_C$, that is,
\begin{equation}
	\eta^C(\cdot) \sim \mathcal{GP} \left( m_C(\cdot), C_C(\cdot, \cdot) + \lambda^2_C  I\right).\nonumber
\end{equation}
We expect that the cheaper simulator's mean is informative for the expensive counterpart and thus, as in \cite{Kennedy2000}, we assume that
\begin{equation}
	\eta^E(\cdot)|\rho, \E\{\eta^C(\cdot)\}, \delta(\cdot) = \rho \E\{ \eta^C(\cdot) \} + \delta(\cdot)\nonumber
\end{equation}
where $\eta^E (\cdot)$ is the expensive stochastic simulator and $\delta(\cdot)$ is a HetGP such that
\begin{align}
	\delta(\cdot)|\lambda_E^2 (\cdot) \sim \mathcal{GP}\left( m_E(\cdot), C_E(\cdot, \cdot) + \lambda^2_E(\cdot)I \right) \nonumber\\
\log \lambda^2_E(\cdot) \sim \mathcal{GP} \left( m_V(\cdot), C_V (\cdot, \cdot) + \lambda_{V}^2 I \right),\nonumber
\end{align}
where the $I$ are identity matrices of appropriate dimensions.
In this formulation, $\rho \in \R$ is a regression parameter and $m_E(\cdot)$, $C_E(\cdot, \cdot)$ are mean and covariance functions for $\delta(\cdot)$. The term $\delta(\cdot)$ serves a dual purpose. Firstly, $\delta(\cdot)$ can be viewed as a discrepancy function; the mean of $\delta(\cdot)$ represents the difference in the mean response of the two simulators, or the loss of accuracy from running cheap simulations (with a large time step/coarse grid). Secondly, $\delta(\cdot)$ describes the stochasticity in the expensive simulator. This is a similar structure to that of Bayesian calibration of deterministic computer models \citep{Ohagan01}, however we do not observe data from a physical system --- but a computer simulator --- and we have noise in both sets of observations.

This joint model for the two simulators allows us to borrow information from the cheaper simulator, but is sufficiently flexible to reject a relationship between the two levels if no such relationship exists. If $\rho = 0$ we recover HetGP.

We express the mean functions in a hierarchical form so that	$m_C(\bx) = h(\bx)^T \bm{\beta}_C$ and $m_E(\bx) = h(\bx)^T \bm{\beta}_E$. We take $h(\cdot)$ to be a set of known, deterministic basis functions. The mean functions have the same form; the particular parameters of these regression functions are allowed to differ.

We will use squared exponential covariance functions so that
\begin{equation*}
  C_{*}(\bx, \bx') =  \sigma^2_{*} \exp\left\{ -(\bx - \bx')^T D^{-1}_{*}(\bx -  \bx')\right\}\nonumber
\end{equation*}
where $* \in \{C, E\}$, $D_{*} = \diag(\theta_{1, *}^2, \ldots, \theta_{K, *}^2)$ is a diagonal matrix containing the correlation lengthscales and $\sigma_*$ are scale parameters of the covariance functions. Note that the choice  of squared exponential covariance function is not a requirement;  the user can specify a different covariance structure as they see fit \citep{Rasmussen2006}.

Since we are only interested in the cheap simulator's mean, we do not consider that it is necessary to estimate a surface for its variance. In fact, the homoscedastic GP is quite good at learning the mean response surface, even in the face of heteroscedasticity (see Fig. $5$ of \cite{Binois2019}).  In our model formulation, $\lambda_{V}$ is a constant nugget for the latent variance of the expensive simulator. Both $\lambda_C$ and $\lambda_V$  smooth the noisy simulator observations. Hence a SML emulator has a similar structure to the standard multilevel emulators presented by \cite{Kennedy2000}, with the addition of a latent variance process ($\lambda_E^2(\cdot)$). We model the log variance as a GP to enforce positivity.

It follows that, conditional on all hyperparameters,\newline $\bm{Y}^T = \left( (\bm{Y}^C)^T, (\bm{Y}^E)^T \right)= (Y^C_1, \ldots, Y^C_{N_C}, Y^E_1, \ldots, Y^E_{N_E})^T$ are multivariate normal where $N_C$ and $N_E$ are the number of runs of the cheap and expensive simulators, respectively. That is,
\begin{equation*}
\begin{pmatrix} \bm{Y}^C \\ \bm{Y}^E \end{pmatrix} \mid \Theta \sim \mathcal{N}_{N_C + N_E} \left\{ \begin{pmatrix} m_C(X^C) \\ \rho m_C(X^E) + m_E(X^E) \end{pmatrix}, \var(\bm{Y}\mid \Theta)\right\}\nonumber
\end{equation*}
where $X^C$ and $X^E$ are sets of input vectors of the cheap and expensive codes, respectively. Details of the design we use are given in \Cref{Sec:design}.

We now derive the covariance matrix of the response $\bm{Y}$ . We write this covariance matrix in block form
\begin{equation}
  \var(\bm{Y} \mid \Theta) = \begin{pmatrix}
  \var(\bm{Y}^C\mid \Theta) & \cov(\bm{Y}^C, \bm{Y}^E\mid \Theta) \\
  \cov(\bm{Y}^E, \bm{Y}^C\mid \Theta) & \var(\bm{Y}^E\mid \Theta)
  \end{pmatrix}.\nonumber
\end{equation}
The auto-covariance of $\bm{Y}^C$ is
\begin{equation*}
\var(\bm{Y}^C\mid \Theta)_{i,j} = \sigma^2_C \exp \left\{ -(\bx^C_i - \bx^C_j)^T D^{-1}_C (\bx^C_i - \bx^C_j)\right\} + \lambda_C^2 \mathbb{I}_{\bx^C_i, \bx^C_j},\nonumber
\end{equation*}
\noindent where $\mathbb{I}_{i, j}$ is an indicator function equal to $1$ when $i=j$ and $0$ otherwise. For the auto-covariance of the expensive simulator, we assume the three summed GPs are all pairwise independent and that the constant variance of the cheap simulator is independent of the variance of the expensive simulator. Further we assume, for $i \neq j$, that
\begin{align*}
\cov(Z^C(\bx_i), \delta(\bx_j)) &= 0\nonumber \\
\cov(Z^C(\bx_i), \lambda^2_E(\bx_j)) &= 0\nonumber \\
\cov(\delta(\bx_i), \lambda^2_E(\bx_j)) &= 0,\nonumber
\end{align*}
where $Z^C(\bx)=\E\{ \eta^C(\bx) \}$. Thus we find that
\begin{align*}
\var(\bm{Y}^E\mid \Theta)_{i,j} & = \cov(Y^E(\bx^E_i), Y^E(\bx^E_j)\mid \Theta)\nonumber\\
& = \rho^2 \sigma_C^2 \exp\left\{ -(\bx^E_i - \bx^E_j)^T D^{-1}_C (\bx^E_i - \bx^E_j) \right\} \nonumber\\
& \hspace{1cm} + \sigma_E^2 \exp\left\{ -(\bx^E_i - \bx^E_j)^T D^{-1}_E (\bx^E_i - \bx^E_j) \right\}  + \lambda^2_E(\bx^E_i)\mathbb{I}_{\bx^E_i, \bx^E_j},\nonumber
\end{align*}
where the $\varepsilon^E_i(\bx_i)$ represents the input dependent noise at $\bx_i$ and $\varepsilon^C_j$ is the (assumed) constant noise exhibited in the cheap simulator at $\bx_j$. Finally, the cross-covariance is given by $\cov(\bm{Y}^C, \bm{Y}^E \mid \Theta)_{i,j}  = \rho C_C(\bx_i, \bx_j).$

Adopting a Gaussian prior for the $\beta$ parameters allows them to be analytically integrated out. For example, if we take
\begin{equation*}
  \begin{pmatrix}
    \bm{\beta}^C \\ \bm{\beta}^E
  \end{pmatrix}\sim \mathcal{N}(\bm{b}, B)\nonumber
\end{equation*}
then we can write $\bm{Y}\mid\Theta_{-\beta}\sim\mathcal{N} \left\{ H\bm{b}, K_0  \right\}$ as the prior for $\bm{Y}$ conditional on the GP covariance matrix where $K_0=\var\{\bm{Y}\mid \Theta\} + HBH^T$ and $H$ is the design matrix. Details of $H$ are discussed in Section~\ref{sec:postpred}.

\subsection{Prior specification}

Since a Bayesian approach to inference is adopted, we assign priors to all GP parameters. We propose that all parameters are assumed independent \textit{a priori} with the following distributions (where the hyperparameters of the prior are chosen by the user),
\begin{align}
\beta_{j, *} &\sim \mathcal{N}(m_{j,*}, s_{j,*}^2) & \theta_{j,*} &\sim  Gamma(a_{j, *}, b_{j, *}) \nonumber\\
\sigma_* &\sim Inv-Gamma(c_{j,*}, d_{j,*}) & \lambda_*^2 &\sim Inv-Gamma(e_{j,*}, f_{j,*})\nonumber \\
\rho &\sim \mathcal{N}(m_{\rho}, s_{\rho}^2),&\nonumber
\end{align}
\noindent where $* \in \{ C, E, V \}$. Note that there is no $\lambda^2_{E}$ since we replace this by a GP to account for heteroscedasticity. For $\beta_{j,*}$ we adopt independent $\mathcal{N}(0,1)$ priors. Because our GP is on the probit scale this prior covers a wide range of observable values; a more diffuse prior (say $s_{j,*}^2=10$) would imply that the simulator output will be very close to either $0$ or $1$ but not between. Our priors on $\theta_{*}$ will be reasonably uninformative, but designed to omit very large lengthscales, therefore we take $a_{j,*} = 2$ and $b_{j,*} = 1$. Fairly weak priors are taken over $\sigma_*$ $c_{j,*}=d_{j,*}=2$ and for $\lambda^2_*$ we have $e_j = f_j=2$. In the prior for $\rho$ we are being quite subjective, we take $m_\rho = 1$ and $s_\rho = 1/3$. This specification expresses the belief that the codes are positively correlated with a high probability; this is a reasonable assertion (recall Figure~\ref{Fig:cheapandexp}). If this belief was not held, then there would be little reason to construct a multilevel emulator. This specification is \textit{our} prior specification. In practice, a user can choose a prior that they see as suitable.
\subsection{Design \label{Sec:design}}
We require a space filling design for both the cheap and expensive versions of the simulator, hence we will appeal to a nested design based on Latin hypercubes. We generate $X^E$ via a maximin Latin hypercube \citep{Mckay1979} (using the \verb|lhs| package in \verb|R|). To generate $X^C$ we make another maximin Latin hypercube and append the two designs together. We run both the cheap and expensive versions of the simulator at $X^E$, but run only the cheap simulator at $X^C$.
\subsection{Posterior predictive distribution of code output}
\label{sec:postpred}
Within our Bayesian approach, MAP estimates will be used to estimate the GP covariance structure. As with HetGP, we integrate out all $\beta$ parameters analytically. MAP estimates are found via a numerical optimisation of the log-posterior (up to an additive constant) using the \verb|optimizing| function from \verb|rstan| \citep{stan}. This is not fully Bayesian, however it is computationally thrifty.

After integrating out the $\beta$ coefficients, we condition on MAP estimates of the remaining parameters to obtain the posterior distribution for $\log \lambda^2_E (X^{\star})$. The posterior at new inputs $X^{\star}$ is Gaussian with mean
\begin{equation*}
m^\star_V (X^{\star}) = H^{\star}_v \bm{b}_V + K_V(X^{\star}, X^E) \big\{ K_V(X^E, X^E) + \lambda_{V}^2 I_E \big\} ^{-1} (\log (\lambda^2_E(X^E)) - H^{\star}_V \bm{b}_V )\nonumber
\end{equation*}
where $K_V(\cdot, \cdot)$ is the same as for HetGP.

\begin{figure}[ht]
\centering
	\includegraphics[width=0.75\textwidth]{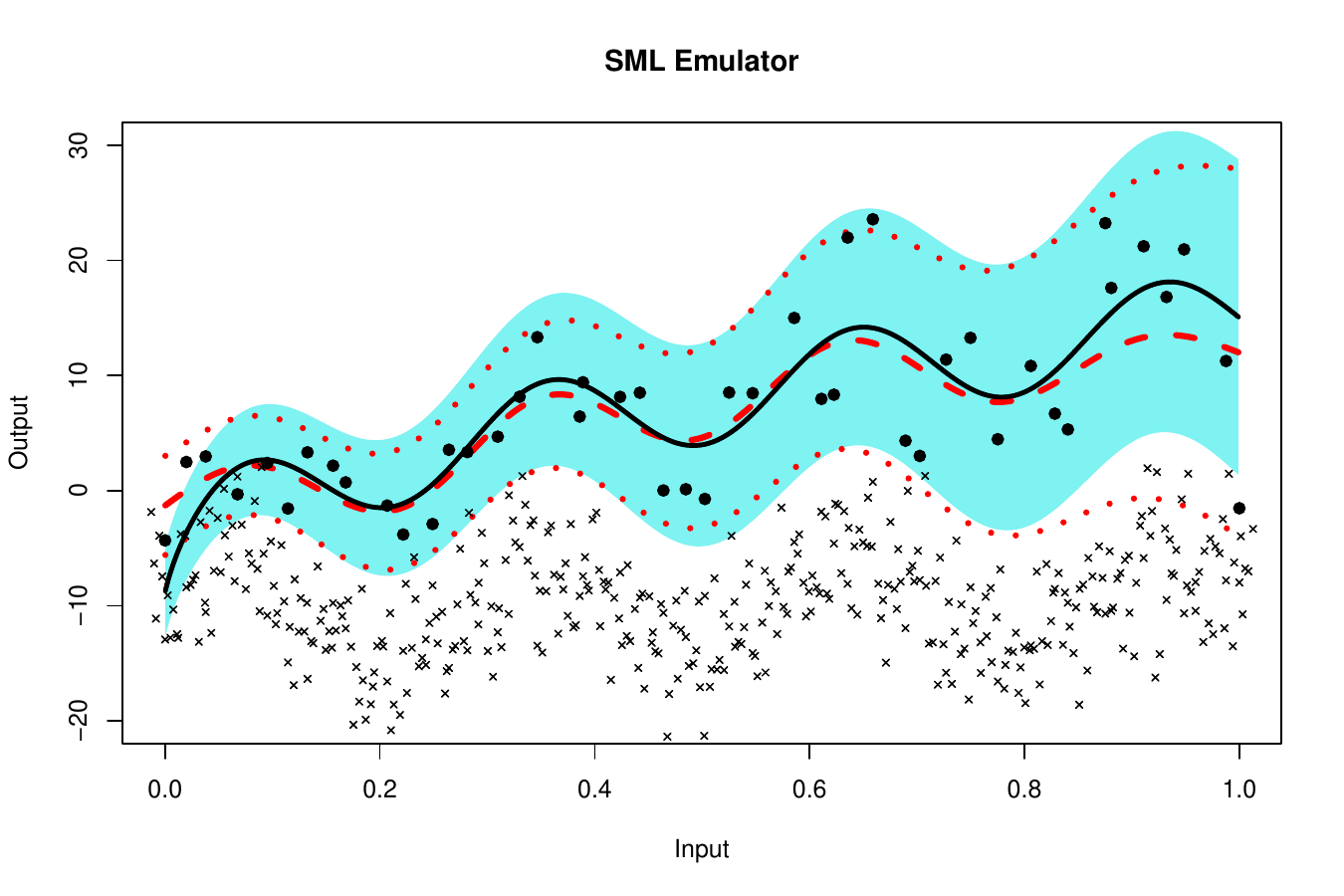}
	\caption{A SML emulator for $\eta(\cdot)$. Expensive runs are black points and cheap runs are black crosses (which are offset by $-10$ to aid visualisation). The true simulator is represented by the black line (mean function) and blue band ($\pm 1.96$ standard deviations). The emulator is represented by the red dashed line (emulator mean) and red dotted lines ($\pm 1.96$ emulator standard deviations).\label{Fig:comparison}}
\end{figure}

Prediction of  $\eta^E(X^{\star})$  is more complex, but is a natural extension of the posterior predictive mean of a two-level code given in \cite{Kennedy2000}. Having observed code outputs $\bm{Y}^C$, $\bm{Y}^E$ at design points $X^C$, $X^E$, our design matrix is
\begin{equation}
H = \begin{pmatrix}
h(\bx^C_1)^T & \bm{0} \\
\vdots & \vdots \\
h(\bx^C_{N_c})^T & \bm{0} \\
 & \\
\rho h(\bx^E_1)^T & h(\bx^E_1)^T \\
\vdots & \vdots \\
\rho h(\bx^E_{N_E})^T & h(\bx^E_{N_E})^T
\end{pmatrix}\nonumber
\end{equation}
and hence the posterior distribution of the output of the expensive simulator at new inputs $X$, conditional on a point estimate of $\Theta_{-\beta}$, is Gaussian with mean
\begin{equation*}
  m^{\star}(X^\star) = h_0(X^\star) \bm{b} + t(X^\star)K_0^{-1}\left( \bm{Y} - H\bm{b} \right).\nonumber
\end{equation*}
If we take $B = \diag(B^C, B^E)$ to be a block diagonal matrix of variance matrices, then the posterior variance, conditional on $\Theta_{-\beta}$, can be expressed as
\begin{align*}
  V^{\star}(X) &= \rho^2 C_c(X^\star,X^\star) + C_E(X^\star,X^\star) + h_0(X^\star)(\rho^2B^C + B^E)h_0(X^\star)^T\\
    &\hspace{1cm}+\lambda^2_E(X^\star)I - t(X)K_0^{-1}t(X)^T,
\end{align*}
where $h_0(X^\star) = ( h(X^\star), h(X^\star))$ and $t(X^\star)=\cov(\eta^E(X^\star), \bm{Y})$.
To get a flavour for SML emulation we have produced an SML emulator in Figure~\ref{Fig:comparison} for the simulator described in Section~\ref{sec:hetgp}. We used $46$ of the runs from the HetGP emulator of Figure~\ref{Fig:toy-HetGP} and replaced them with $400$ runs from a `cheap' simulator $\eta^C(x) = 4 \sin (7\pi x) + 4\varepsilon$ with $\varepsilon \sim N(0,1)$ and $x \in [0,1]$.  The cheap points have a similarly shaped mean function to the expensive points. This information is utilised by the SML emulator to provide an emulator which closely mimics $\eta(\cdot)$.

\section{Stochastic multilevel emulation of the Athena simulator}
\label{sec:SML-athena}

We return to the motivating example for the SML emulator; the Athena simulator. Recall, from Section~\ref{sec:athena}, that Athena is a large point-process model. Simulations are implemented via MATLAB with a large number of inputs. Many inputs are parameters of lifetime distributions of components in wind turbines, but others, for instance, relate to the availability of repair equipment. These additional inputs are not considered here; we are interested in the component reliabilities which are most critical to offshore wind farm availability. Specifically, we focus on inputs which are the times of onset of degradation of the nine key wind farm components mentioned in Section~\ref{sec:athena}, and indexed as follows: 1. gearbox, 2. generator, 3. frequency converter, 4. transformer, 5. main shaft bearing, 6. the blades, 7. tower, 8. foundations and 9. the catch all.

\subsection{Emulator construction}
\label{sec:em-con}

We construct emulators over a $9$ dimensional input space.  We vary each input over the range $[0.1, 5]$ (years). The Athena simulator is flexible enough to specify unique parameters for every subassembly in every turbine. We give the same parameter values to each subassembly of a given type and allow different types of subassembly to have different parameters. For example, all gearboxes could have a time to degradation of $1$ year whereas all generators could have a time to degradation of $3.2$ years.

Design points are chosen via the structure described in \Cref{Sec:design}. To construct the HetGP emulator we ran the Athena simulator at $100$ design points. The cheap runs of the simulator were fast enough that we could trade just $5$ expensive design points for $295$ cheap runs. We used basis functions $h(\bx) = (1, \log(\bx))$ for the mean functions of the mean response. We arrived at this selection to reflect a prior belief that the mean availability would flatten off at larger values of $x_i$. The covariance function assumes standardised inputs, $x_i^*$. Standardisation is achieved by subtracting the sample means and then dividing by the sample standard deviations (of the expensive training data). The latent variance GP has mean function $m_V(\bx) = (1, \bx^*)\bm{\beta}_V$ and again, the covariance function assumes standardised inputs.
\begin{figure}
	\centering
	\includegraphics[width=0.7\textwidth]{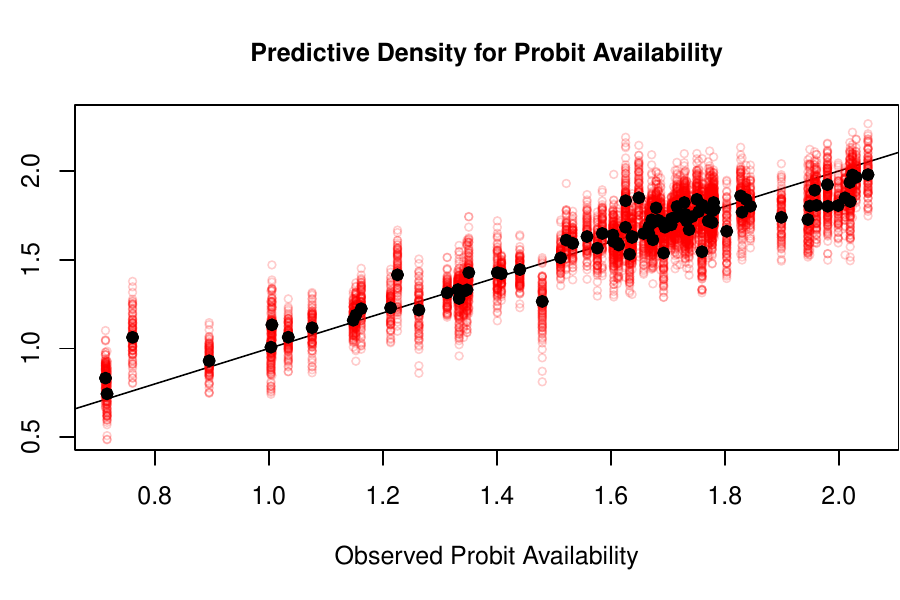}
\caption{Observed probit availability (training data) plotted against emulator mean predictions (black dots) and $100$ realisations from the emulator at each point (translucent red circles).\label{Fig:within-sample}}
\end{figure}
\Cref{Fig:within-sample} shows the emulator predictions of the training data (probit scale) with realisations from $\mathcal{N}\{m^{\star}(\bx), V^{\star}(\bx)\}$ around each prediction. Large deviations from the unit diagonal are typically accompanied by a more diffuse predictive distribution; the emulator is giving larger variance to the points which are far away from the mean. We also see that the observed probit availabilities are mostly in the region of $0.5$ -- $2$ (availabilities in the region of around $0.76$ -- $0.98$). The full range of observed availabilities is $(0.762, 0.980)$; the vast majority of realisations from the emulator agree with this range.

\subsection{Emulator performance comparison}

To judge relative performance of each emulator we propose using two metrics. The first is root mean squared error (RMSE), comparing the unseen simulator realisations to the emulator predictive mean. The second performance measure is a proper scoring rule; we use the scoring rule given in Equation~($27$) of \cite{Gneiting2007}.  This scoring rule has also been used in the emulator literature \citep{Binois2018, Baker2020a}. A larger score suggests better fit.

Using $100$ independently generated validation data points, the RMSE (on the probit scale) for HetGP was $0.181$, whereas SML achieved an RMSE of $0.156$. The score for HetGP was $238$ and for SML the score was $254$ (probit scale). Since we transformed the availability to construct emulators on an unbounded space, we should also check how predictions perform on the $[0,1]$ scale. Using an inverse-probit transformation on the mean function provides a sensible point estimate of availability. Comparing the MSE on the original scale  we observe an RMSE of $0.0272$ for HetGP, and under SML this is reduced to $0.0198$. Hence, SML achieves better RMSE and score here than HetGP for the Athena model, suggesting it is a better emulator. Further, our MAP estimate of $\rho$ is $\hat{\rho} = 0.51$. This suggests a moderate correlation between the two versions of Athena. The additional information extracted from cheap simulations has improved our emulation with little computational cost. It took $5.7$ seconds to fit HetGP and $29.6$ seconds to fit SML on a laptop with $4 \times 2.40\,\text{GHz}$ processors and $8\,\text{GB}$ RAM. Although SML took more time to fit, in real terms this is about $30$ seconds of computation time -- less than a single  expensive run of Athena. Both timings are for a total of $3$ fits of the emulator. We performed $3$ fits to prevent choosing a local mode as the MAP estimate.

\subsection{Emulator validation}

To validate the emulators, we will implement some graphical diagnostics proposed by \citet{Bastos09}. Since we model the (transformed) simulator outputs by a Gaussian process, the Cholesky errors (CEs) should form a random sample from a $\mathcal{N}(0, 1)$ distribution (approximately). If the posterior mean and variance are well suited to the simulator, the validation data should lie in a horizontal band, centred at $0$, with approximately $95\%$ of points in the interval $(-1.96, 1.96)$. We also compare empirical quantiles of the CEs against theoretical quantiles -- we do this via coverage plots which compare the proportion of CEs in the $100(1-\alpha)\%$ prediction interval against the expected proportion.
\begin{figure}[!ht]
    \centering
       \includegraphics[width = 0.8\textwidth]{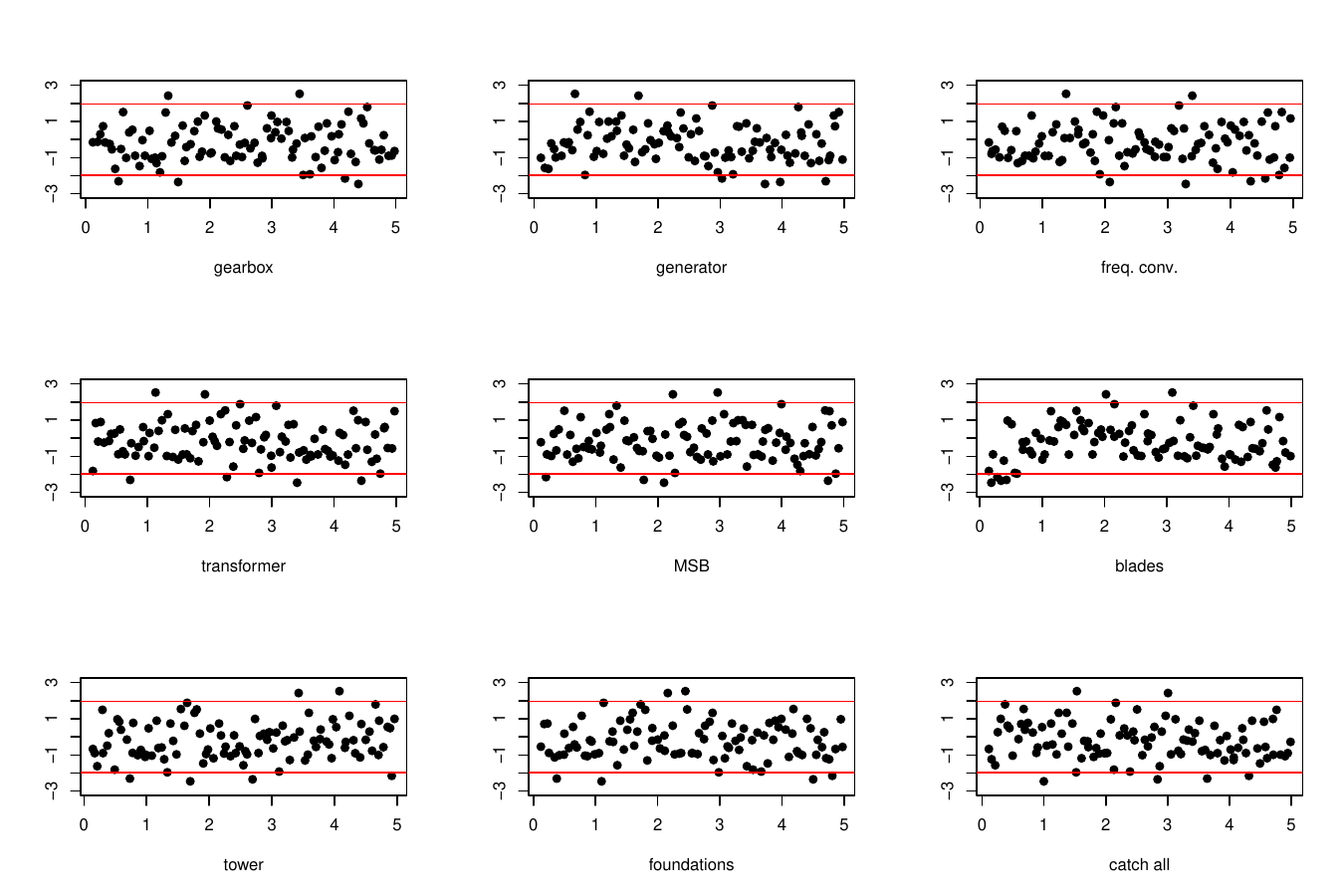}
       \caption{Cholesky errors for HetGP, based on $100$ ``unseen'' validation points. Orange lines are at $\pm 1.96$.\label{Fig:het-resids}}
\end{figure}
\begin{figure}[!ht]
    \centering
\includegraphics[width = 0.8\textwidth]{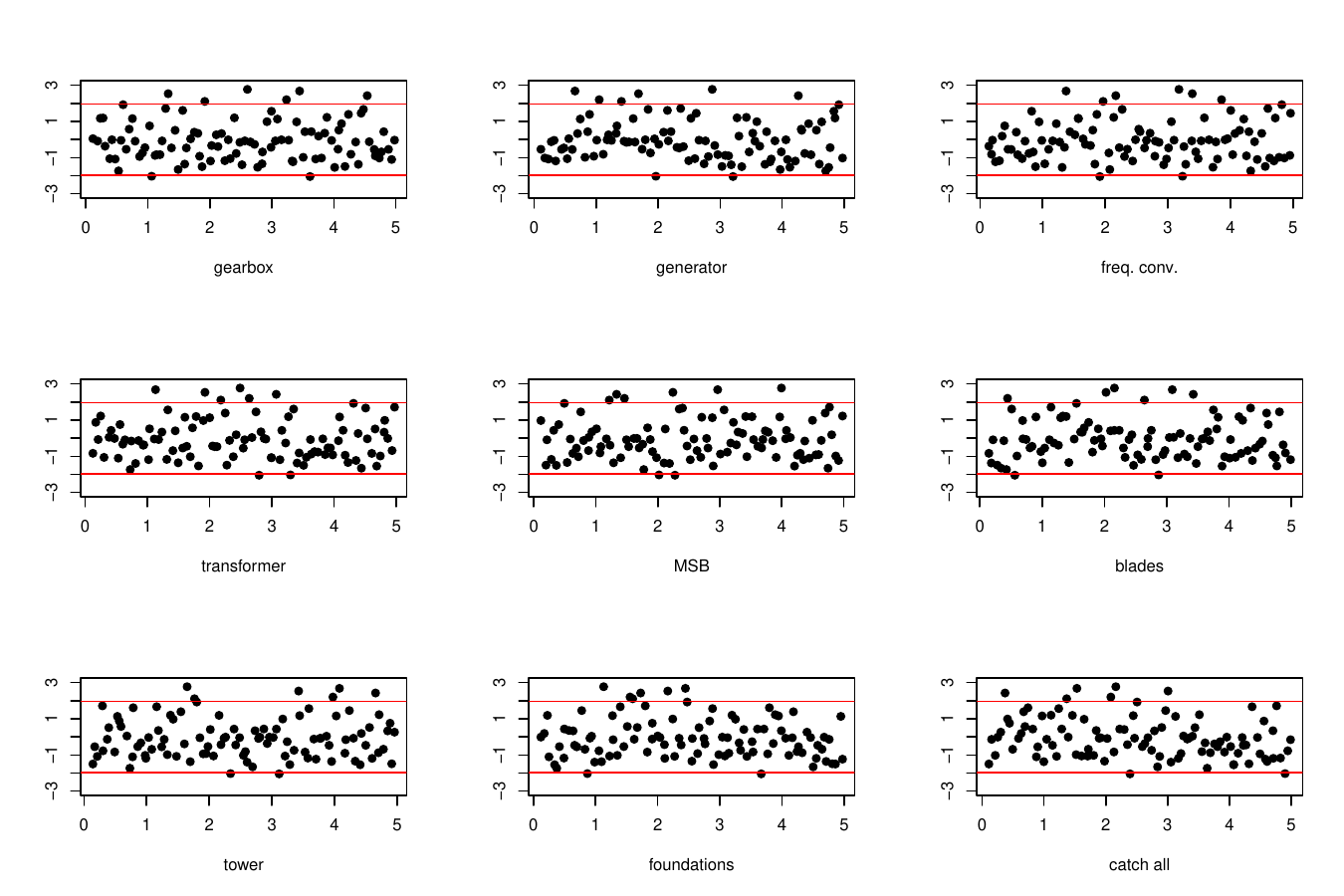}
\caption{Cholesky errors for SML emulation, based on $100$ ``unseen'' validation points. Orange lines are at $\pm 1.96$.\label{Fig:sml-resids}}
\end{figure}
\begin{figure}[!ht]
    \centering
\includegraphics[width = 0.7\textwidth]{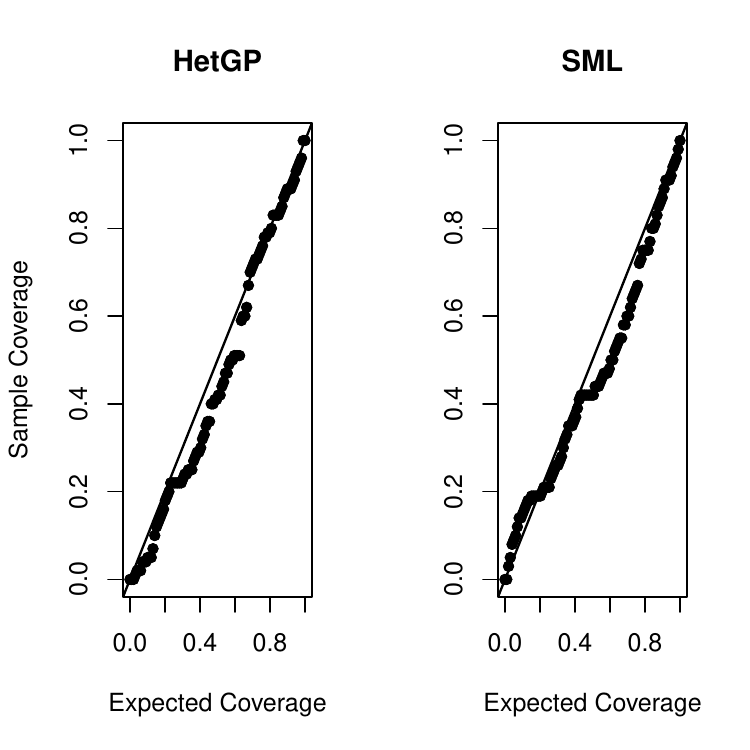}
    \caption{Out of sample coverage plots (black dots), using the Cholesky errors of the ``unseen'' validation data. Black lines represent the unit diagonal. \label{Fig:coverage}}
\end{figure}
In Figure~\ref{Fig:het-resids} the CEs for HetGP have a distinct pattern when plotted against $x_6$, whereas for SML in Figure~\ref{Fig:sml-resids} the points appear to be closer to a random $\mathcal{N}(0,1)$ sample.  The coverage plots in Figure~\ref{Fig:coverage} suggest that for both emulators the coverage is reasonably well calibrated.

\section{Probabilistic sensitivity analysis of Athena}
\label{sec:PSA}

We now use emulators to perform efficient PSA to deduce which of the inputs are the ``driving force'' of the output uncertainty. We also want to understand how much uncertainty is induced by the stochastic nature of Athena. The sensitivity analysis we perform is on the probit-availability scale (the scale the emulator was constructed on). We use the approach of \citet{Marrel2012} to perform PSA, which we outline below.
\subsection{PSA for stochastic simulators \citep{Marrel2012}}
\label{sec:stochPSA}
Performing PSA when the model is stochastic is broadly the same as standard techniques,such as those in \citet{Oakley04}. The addition introduced by \citet{Marrel2012} is to think of the seed as an (unobserved) variable which can be incorporated into the functional ANOVA decomposition. That is, we should think of $\eta(\bx)$ as a function of $\bx$, the inputs, and $x_\varepsilon$, the seed rather than just the inputs alone. The uncertainty  $x_\varepsilon$ is then the uncertainty induced by stochasticity. It is also useful to think of the stochastic computer model as a mean function $Y_m(\bx) = \E\{\eta(\bx)|\bx\}$ and a dispersion function $Y_d(\bx) = \var\{ \eta(\bx)| \bx \}$. Our GP assumptions make higher order moments redundant. Then the total uncertainty in the stochastic computer model output is, by the total variance formula,
\begin{equation*}
V = \var(Y) = \var\{Y_m(\bx) \} + \E\{ Y_d(\bx) \},\nonumber
\end{equation*}
where the expectations and variances are taken over $\bx$. The mean response has an ANOVA decomposition into main effects and interactions,
\begin{equation*}
Y_m(\bx) = f_0 + \sum_i f_i(x_i) + \sum_{i<j} f_{ij}(x_i, x_j) + \sum_{i<j<k} f_{ijk}(x_i, x_j, x_k) + \cdots + f_{1\ldots K} (\bx),\nonumber
\end{equation*}
where $f_0 = \E \{ Y_m(\bx)\} $ is the expected simulator output, $f_{ij}(x_i, x_j)$ is the first order interaction between variables $i$ and $j$, $f_{ijk}(x_i, x_j, x_k)$ denotes the interaction between variables $i$, $j$ and $k$ and so on. We then compute the main effects by
\begin{equation*}
f_i(x_i) = \E_{x_{-i}} \{ Y_m(\bx) | x_i \} - f_0.\nonumber
\end{equation*}
The observed response (accounting for stochasticity) is
\begin{equation}
\eta(\bx) = Y_m(\bx) + f_{\varepsilon}(\bx) + \sum_{  J  \subseteq \{1, \ldots , K\} } f_{ \varepsilon J}(\bx), \label{Eq:observed}
\end{equation}
where $f_\varepsilon (\bx)$ is the main effect of the seed and $f_{\varepsilon J}(\bx)$ is the interaction between the seed and the variables attributed to subset $J$. The main effects and interactions determine how much of the uncertainty in $Y_m(\bx)$ is attributed to a particular subset of the inputs $J \subseteq \{1, 2, \ldots, K\} $,
\begin{equation*}
	V_J = \sum_{J' \subseteq J} \var \{ f_{J'} (\bx_{J'}) \}.\nonumber
\end{equation*}
Normalising these variances by $V$ gives us a scaled quantity $S_J = V_J / V \in [0,1]$ which is the proportion of variance in $Y$ induced by the uncertainty in $\bx_J$. These $S_J$ are often called Sobol' sensitivity indices. However, in the stochastic setting, $S = \sum_{i} S_i + \sum_{i<j} S_{ij} + \ldots + S_{1\ldots K}  =\var \{ Y_m(\bx)\} / V  < 1$. The remaining uncertainty is accounted for by $S_{T_\varepsilon} =  \E \{Y_d(\bx)\}/V$; the total uncertainty induced by the random seed or stochasticity.

The analysis can be performed for $\log \lambda^2(\bx)$ with sensitivity indices denoted $S_{*}^{\lambda}$. Hence $\log Y_d(\bx)=\log \lambda^2(\bx)$ has ANOVA decomposition
\begin{equation*}
  \log Y_d(\bx)= f_0^{\lambda} + \sum_{J \subseteq \{1, 2, \ldots, K\}} f^{\lambda}_J(\bx_J) + f_{\varepsilon}^{\lambda}(\bx)+\sum_{J \subseteq \{1, 2, \ldots, K\}} f^{\lambda}_{\varepsilon J}(\bx_J).\nonumber
\end{equation*}
 Since $\log \lambda^2(\bx)$ has a constant nugget, $S_{T_{\varepsilon}} =  \lambda^2_V/V_\lambda$ and $S_{\varepsilon J} = 0$ for non-empty $J$.
\subsection{Application of stochastic PSA to Athena}
\begin{figure}[ht]
     \centering
   \includegraphics[width=0.75\textwidth]{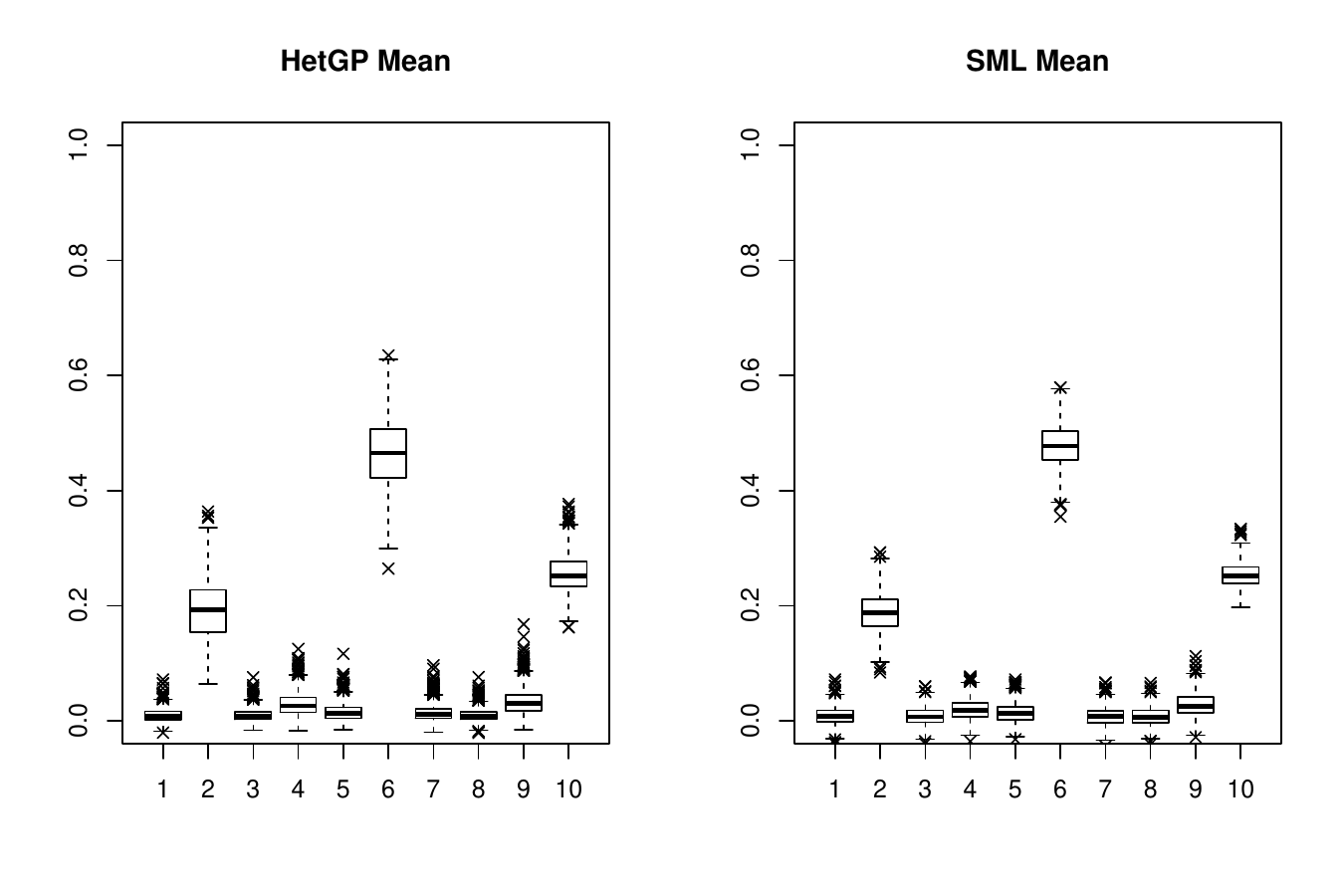}
    \caption{Boxplots representing the posterior distribution of $S_i$; $i=1$, $2$, \ldots, $9$. The $10^{th}$ index corresponds to $S_{T_\varepsilon}$. Left hand plot corresponds to HetGP; right hand to SML.}  \label{Fig:si-mean}
\end{figure}

\begin{figure}[ht]
      \centering
   \includegraphics[width=0.75\textwidth]{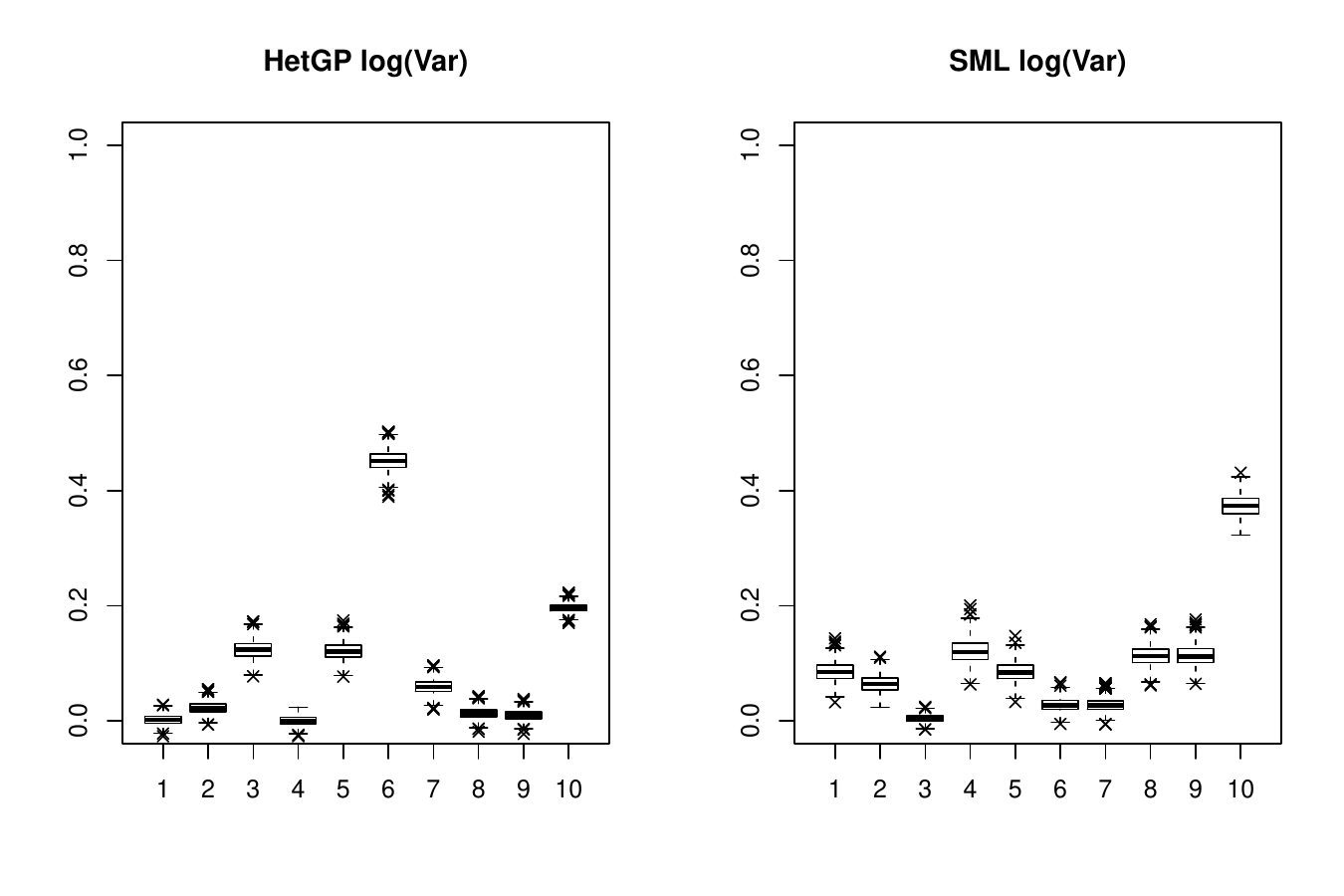}
   \caption{Boxplots representing the posterior distribution of $S_i^{\lambda}$; $i=1$, $2$, \ldots, $9$. The $10^{th}$ index corresponds to $S_{T_\varepsilon}$. Left hand plot corresponds to HetGP; right hand to SML.}
   \label{Fig:si-var}
 \end{figure}

To estimate all the above quantities from Section~\ref{sec:stochPSA}, we replace the Athena simulator, $\eta(\cdot)$, with an emulator. We compare the estimation under HetGP and SML. All relevant quantities are estimated by Monte Carlo simulation to compute the expectations and variances with respect to $\bx$, conditional on all GP parameters. It is common in PSA to give a simple probability distribution to the inputs of interest \citep{Kennedy2006,Saisana2005,Overstall2016}.  Our parameters are each assumed to follow $U(0.1, 5)$ distributions, covering the range for which the emulators were constructed; see Section~\ref{sec:em-con}. Our estimation approach is Bayesian; we draw $1000$ different values of the $\beta$ parameters from the posterior distribution and then for each draw compute Sobol' sensitivity indices based on Latin hypercube samples of size $N=10^4$. Boxplots of first order indices based on both HetGP and SML are given in \Cref{Fig:si-mean}. In both cases, $\sum_{i=1}^9S_i + S_{T_\varepsilon}\approx 1$ suggesting Athena is approximately additive in the inputs.
 Estimated first order sensitivity indices for the mean give us more-or-less the same interpretation about the Athena simulator. We see in both cases that $S_6  > S_{T_\varepsilon} > S_2$ are the three most important indices, the rest have mean value comfortably under $5\%$. Since $S_{T_\varepsilon}$ is clearly larger than all but one first order effect, this suggests the stochasticity in the Athena simulator is an important part of the model (randomness contributes roughly as much uncertainty as $x_2$).
  The estimates of $S^{\lambda}_i$ are very different under the two approaches.
 Observing \Cref{Fig:si-var} we can see that the HetGP estimate of $S^{\lambda}_6$ is very large (around $50\%$) whereas under SML the estimate is less than $10\%$. We suspect that HetGP is interpreting a large proportion of the systematic variation due to $x_6$ as noise whereas SML gives a much improved interpretation of the variation.
 \begin{figure}[ht]
   \centering
   \includegraphics[width=0.75\textwidth]{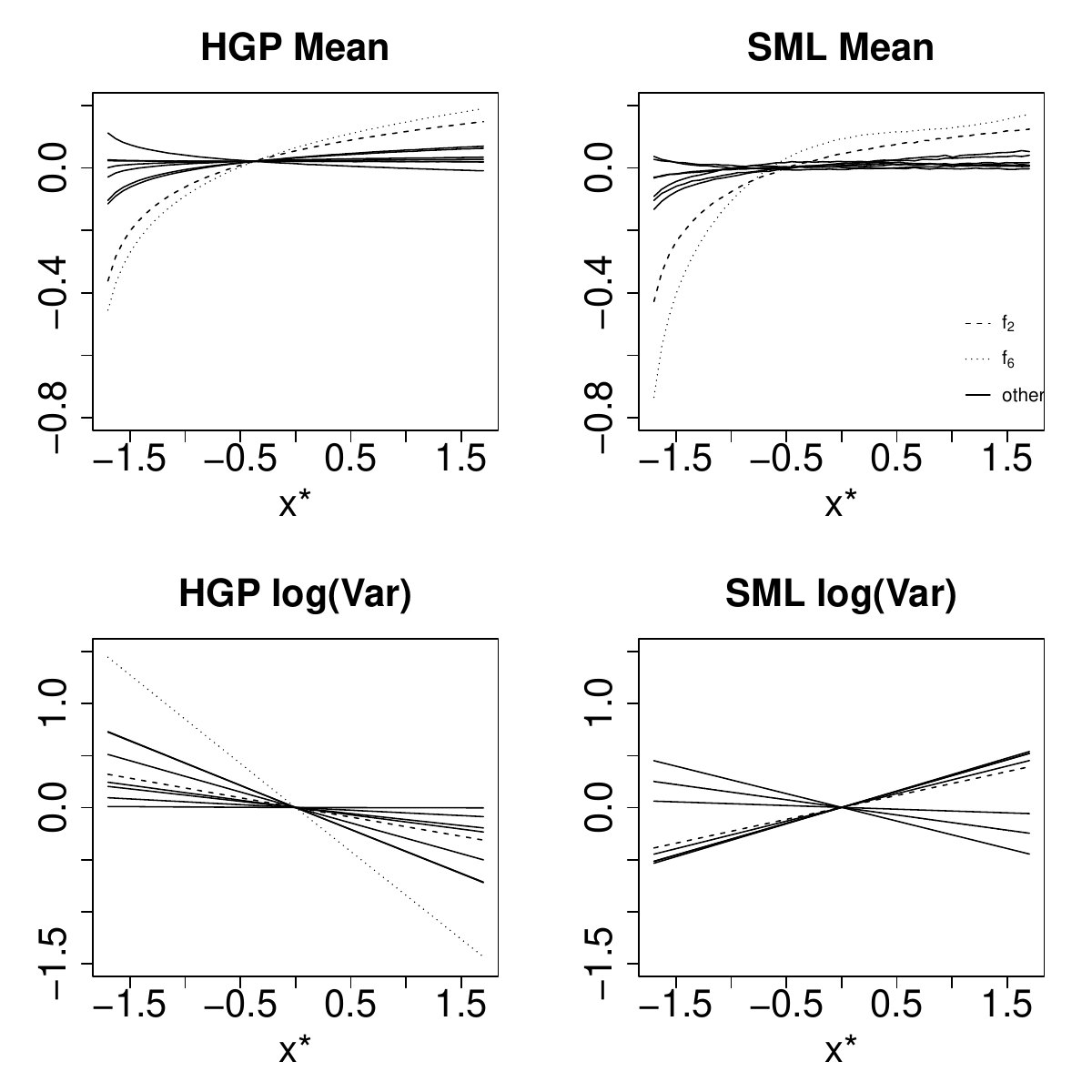}
     \caption{Main effect plots under HetGP (left) and SML (right). Top plots correspond to the mean surface and bottom plots to the log variance surface. The dashed lines correspond to $f_2(x_2)$ and $f_2^{\lambda}(x_2)$, dotted lines to $f_6(x_6)$ and $f_6^{\lambda}(x_6)$. The solid lines represent all other main effects. The $x$ axis is on the standardised scale that the emulators were fitted on. Note that the scale of the $y$ axis on the mean plot differs from that of the variance plot.}
  \label{Fig:main-eff-all}
 \end{figure}
  We see that qualitatively, the main effect plots (\Cref{Fig:main-eff-all}) agree with the estimated values of $S_i$ and $S^{\lambda}_i$. That is, an input with high $S_i$ exhibits a large range in the main effect plot. The main effect plots for the mean are quite similar with the exception of $f_6(x_6)$. Under SML $f_6(x_6)$ has a much larger range and the shape under HetGP is closer to $\log(x_6)$ --- the chosen basis function. This suggests that SML is borrowing information from the cheap simulator to inform the mean response of the expensive simulator and marries up with the (lack of) structure in the residual plots seen earlier (\Cref{Fig:het-resids}, \Cref{Fig:sml-resids}). The main effects for the log variance are quite different under the two emulators. Under HetGP, $x_6$ is highly influential for the log variance whereas $x_6$ is much less influential under the SML emulator. A stark difference is that the slopes $f_6^{ \lambda }(x_6)$ are of opposite sign. The slope of the $f_6^{ \lambda }(x_6)$ under SML is positive which matches up with \Cref{Fig:log-vars}. We believe this is due to SML resolving a kind of weak identifiability issue; when insufficient data is fed to a HetGP emulator, it will frequently model systematic variation as noise. This was also seen in the toy example (\Cref{Fig:toy-HetGP}).

We have now performed the necessary analyses to set up a future elicitation procedure. Using the results from either the SML or HetGP emulator the largest contributions to output uncertainty were the failures of the blades ($x_6$) and generator ($x_2$). An equi-tailed $95\%$ credible interval (computed via SML) for $S_2 + S_6$ is $(59,73)\%$ of input uncertainty. Our planning of the elicitation of parameters would focus mainly on these two parameters, since they jointly contribute to over half of output uncertainty. The other inputs would not be completely neglected since they contribute roughly equal amounts of uncertainty to the log variance. Without an emulator this sensitivity analysis would have taken many months of CPU time. Our SML emulator allowed us to further reduce the amount of time required to construct an adequate emulator by exploiting a computationally cheaper version of the Athena simulator.
\section{Conclusions}
\label{sec:conc}
We have introduced a stochastic multilevel emulator, which adopted elements of (i) the autoregressive structure from \cite{Kennedy2000} to construct a more accurate mean function and (ii) the latent variance structure of HetGP to account for a heteroscedastic computer model. This structure allowed us to link together two versions of the Athena simulator to perform accurate and efficient sensitivity analysis. The easy to generate training data allowed us to build an adequate emulator without having to spend many days generating training data.

It would be interesting to see, from a methodological point of view, how the SML emulator could be improved. One idea would be to implement a sequential design rule similar to that of \citet{Gratiet2015}, that is, minimising some design criterion such as integrated mean squared prediction error. Another idea would be to use a preliminary round of simulations to see where cheap simulations might be most beneficial. For example, it might be beneficial to place more cheap points where the two levels agree most and then retain the expensive simulation budget for areas where the two levels disagree. It would also be interesting to see if replicates could improve this type of emulator in the same way that replicates benefit HetGP. Replication could be especially beneficial in the cheap simulator; this would help to reduce the size of computational overheads of a large design matrix since inference is $\mathcal{O}(N^3)$ for HetGP and $\mathcal{O}\left((N_C+N_E)^3\right)$ for SML. Prediction is $\mathcal{O}\left(N^2\right)$ for HetGP and $\mathcal{O}\left((N_C+N_E)^2\right)$ for SML. Another possibility would be to link the variance of the two simulators; we chose not to do this as it would involve linking two latent variance processes and would involve inversion of a large matrix, increasing the computational cost of inference and prediction.

A detailed probability elicitation can be very useful in the event of limited data, such as our wind farm setting. Although there is a large amount of data from existing wind farms, this is not directly relevant to a future wind farm so should not be blindly applied to a future wind farm. The probability elicitation should focus on the most important parameters and, in this paper, these were found to be the times to degradation of the generator and the blades.

Ultimately, we wish to utilise an emulator of the Athena simulator in a Bayesian decision analysis to allow a decision maker to answer questions concerning the design and maintenance strategy of large offshore wind farms. This would involve eliciting a utility function over availability and other features such as, but not limited to, the monetary cost of particular turbine components.
\section*{Code}
\texttt{R} code and data to fit, and validate, HetGP and SML emulators for the Athena example are available from \texttt{github.com/jcken95/sml-athena}.
\section*{Acknowledgements}
We would like to thank two anonymous referees and the associate editor for insightful comments which have considerably improved the manuscript. We would also like to express further gratitude to the Engineering and Physical Sciences Research Council (EPSRC) for JCK's studentship, and the EPSRC UK Centre for Energy Systems Integration, grant number $EP/P001173/1$ for further financial support. Finally, we are grateful to Professor Tim Bedford and Professor Lesley Walls of the University of Strathclyde for useful discussions about the Athena simulator.
\bibliography{references}	

\begin{thebibliography}{1}

\bibitem{Binois2019}
Mickaël Binois, Jiangeng Huang, Robert~B. Gramacy, and Mike Ludkovski.
\newblock Replication or exploration? sequential design for stochastic
  simulation experiments.
\newblock {\em Technometrics}, 61(1):7--23, 2019.

\end{thebibliography}


\begin{thebibliography}{xx}

\harvarditem[Andrianakis et~al.]{Andrianakis, Vernon, McCreesh, McKinley,
  Oakley, Nsubuga, Goldstein \harvardand\ White}{2017}{Andrianakis2017}
Andrianakis, I., Vernon, I., McCreesh, N., McKinley, T., Oakley, J., Nsubuga,
  R., Goldstein, M. \harvardand\ White, R.  \harvardyearleft
  2017\harvardyearright , `History matching of a complex epidemiological model
  of human immunodeficiency virus transmission by using variance emulation',
  {\em Journal of the Royal Statistical Society. Series C, Applied Statistics}
  {\bf 66}(4),~717.

\harvarditem[Ankenman et~al.]{Ankenman, Nelson \harvardand\
  Staum}{2010}{Akenman2010}
Ankenman, B., Nelson, B.~L. \harvardand\ Staum, J.  \harvardyearleft
  2010\harvardyearright , `Stochastic {K}riging for simulation metamodeling',
  {\em Operations Research} {\bf 58}(2),~371--382.

\harvarditem[Astfalck et~al.]{Astfalck, Cripps, Gosling \harvardand\
  Milne}{2019}{Astfalck19}
Astfalck, L., Cripps, E., Gosling, J. \harvardand\ Milne, I.  \harvardyearleft
  2019\harvardyearright , `Emulation of vessel motion simulators for
  computationally efficient uncertainty quantification', {\em Ocean
  Engineering} {\bf 172},~726--736.

\harvarditem{Baker, Barbillon, Fadikar, Gramacy, Herbei, Higdon, Huang,
  Johnson, Ma, Mondal, Pires, Sacks \harvardand\ Sokolov}{2020}{Baker2020}
Baker, E., Barbillon, P., Fadikar, A., Gramacy, R.~B., Herbei, R., Higdon, D.,
  Huang, J., Johnson, L.~R., Ma, P., Mondal, A., Pires, B., Sacks, J.
  \harvardand\ Sokolov, V.  \harvardyearleft 2020\harvardyearright ,
  `Stochastic simulators: An overview with opportunities', {\em arXiv preprint
  arXiv:2002.01321} .

\harvarditem{Baker, Challenor \harvardand\ Eames}{2020}{Baker2020a}
Baker, E., Challenor, P. \harvardand\ Eames, M.  \harvardyearleft
  2020\harvardyearright , `Predicting the output from a stochastic computer
  model when a deterministic approximation is available', {\em Journal of
  Computational and Graphical Statistics} pp.~1--12.

\harvarditem{Bastos \harvardand\ O'Hagan}{2009}{Bastos09}
Bastos, L.~S. \harvardand\ O'Hagan, A.  \harvardyearleft 2009\harvardyearright
  , `Diagnostics for {G}aussian process emulators', {\em Technometrics} {\bf
  51}(4),~425--438.

\harvarditem[Becker et~al.]{Becker, Oakley, Surace, Gili, Rowson \harvardand\
  Worden}{2012}{Becker2012}
Becker, W., Oakley, J., Surace, C., Gili, P., Rowson, J. \harvardand\ Worden,
  K.  \harvardyearleft 2012\harvardyearright , `Bayesian sensitivity analysis
  of a nonlinear finite element model', {\em Mechanical Systems and Signal
  Processing} {\bf 32},~18--31.

\harvarditem{Binois \harvardand\ Gramacy}{2019}{hetGP}
Binois, M. \harvardand\ Gramacy, R.~B.  \harvardyearleft 2019\harvardyearright
  , {\em hetGP: Heteroskedastic {G}aussian Process Modeling and Design under
  Replication}.
\newblock R package version 1.1.1.

\harvarditem[Binois et~al.]{Binois, Gramacy \harvardand\
  Ludkovski}{2018}{Binois2018}
Binois, M., Gramacy, R.~B. \harvardand\ Ludkovski, M.  \harvardyearleft
  2018\harvardyearright , `Practical heteroscedastic {G}aussian process
  modeling for large simulation experiments', {\em Journal of Computational and
  Graphical Statistics} {\bf 27}(4),~808--821.

\harvarditem[Binois et~al.]{Binois, Huang, Gramacy \harvardand\
  Ludkovski}{2019}{Binois2019}
Binois, M., Huang, J., Gramacy, R.~B. \harvardand\ Ludkovski, M.
  \harvardyearleft 2019\harvardyearright , `Replication or exploration?
  sequential design for stochastic simulation experiments', {\em Technometrics}
  {\bf 61}(1),~7--23.

\harvarditem[Boys et~al.]{Boys, Ainsworth \harvardand\
  Gillespie}{2018}{Boys2018}
Boys, R.~J., Ainsworth, H.~F. \harvardand\ Gillespie, C.~S.  \harvardyearleft
  2018\harvardyearright , `Bayesian inference for a partially observed
  birth--death process using data on proportions', {\em Australian \& New
  Zealand Journal of Statistics} {\bf 60}(2),~157--173.

\harvarditem[Carroll et~al.]{Carroll, McDonald \harvardand\
  McMillan}{2016}{Carroll2016}
Carroll, J., McDonald, A. \harvardand\ McMillan, D.  \harvardyearleft
  2016\harvardyearright , `Failure rate, repair time and unscheduled {O}\&{M}
  cost analysis of offshore wind turbines', {\em Wind Energy} {\bf
  19}(6),~1107--1119.

\harvarditem[Forrester et~al.]{Forrester, S{\'o}bester \harvardand\
  Keane}{2007}{Forrester2007}
Forrester, A.~I., S{\'o}bester, A. \harvardand\ Keane, A.~J.  \harvardyearleft
  2007\harvardyearright , `Multi-fidelity optimization via surrogate
  modelling', {\em Proceedings of the Royal Society A: Mathematical, Physical
  and Engineering Sciences} {\bf 463}(2088),~3251--3269.

\harvarditem[Fricker et~al.]{Fricker, Oakley, Sims \harvardand\
  Worden}{2011}{Fricker2011}
Fricker, T.~E., Oakley, J.~E., Sims, N.~D. \harvardand\ Worden, K.
  \harvardyearleft 2011\harvardyearright , `Probabilistic uncertainty analysis
  of an {FRF} of a structure using a {G}aussian process emulator', {\em
  Mechanical Systems and Signal Processing} {\bf 25}(8),~2962--2975.

\harvarditem{Gneiting \harvardand\ Raftery}{2007}{Gneiting2007}
Gneiting, T. \harvardand\ Raftery, A.~E.  \harvardyearleft
  2007\harvardyearright , `Strictly proper scoring rules, prediction, and
  estimation', {\em Journal of the American Statistical Association} {\bf
  102}(477),~359--378.

\harvarditem[Goldberg et~al.]{Goldberg, Williams \harvardand\
  Bishop}{1998}{Goldberg1998}
Goldberg, P.~W., Williams, C.~K. \harvardand\ Bishop, C.~M.  \harvardyearleft
  1998\harvardyearright , Regression with input-dependent noise: A {G}aussian
  process treatment, {\em in} `Advances in {N}eural {I}nformation {P}rocessing
  {S}ystems', pp.~493--499.

\harvarditem{Goldstein \harvardand\ Rougier}{2006}{Goldstein2006}
Goldstein, M. \harvardand\ Rougier, J.  \harvardyearleft 2006\harvardyearright
  , `Bayes linear calibrated prediction for complex systems', {\em Journal of
  the American Statistical Association} {\bf 101}(475),~1132--1143.

\harvarditem{Gramacy}{2020}{Gramacy2020surrogates}
Gramacy, R.~B.  \harvardyearleft 2020\harvardyearright , {\em Surrogates:
  {G}aussian Process Modeling, Design and Optimization for the Applied
  Sciences}, Chapman Hall/CRC, Boca Raton, Florida.
\newline\harvardurl{http://bobby.gramacy.com/surrogates/}

\harvarditem[Harvey et~al.]{Harvey, Huntley, Dacre, Goldstein, Thomson
  \harvardand\ Webster}{2018}{Harvey2018}
Harvey, N.~J., Huntley, N., Dacre, H.~F., Goldstein, M., Thomson, D.
  \harvardand\ Webster, H.  \harvardyearleft 2018\harvardyearright ,
  `Multi-level emulation of a volcanic ash transport and dispersion model to
  quantify sensitivity to uncertain parameters', {\em Natural Hazards and Earth
  System Sciences} {\bf 18}(1),~41--63.

\harvarditem[Henderson et~al.]{Henderson, Boys, Krishnan, Lawless \harvardand\
  Wilkinson}{2009}{Henderson09}
Henderson, D.~A., Boys, R.~J., Krishnan, K.~J., Lawless, C. \harvardand\
  Wilkinson, D.~J.  \harvardyearleft 2009\harvardyearright , `Bayesian
  emulation and calibration of a stochastic computer model of mitochondrial
  {DNA} deletions in substantia nigra neurons', {\em Journal of the American
  Statistical Association} {\bf 104}(485),~76--87.

\harvarditem{Hobley}{2019}{Hobley2019}
Hobley, A.  \harvardyearleft 2019\harvardyearright , `Will gas be gone in the
  {United Kingdom (UK)} by 2050? {An} impact assessment of urban heat
  decarbonisation and low emission vehicle uptake on future {UK} energy system
  scenarios', {\em {Renewable Energy}} {\bf 142},~695--705.

\harvarditem[Kennedy et~al.]{Kennedy, Anderson, Conti \harvardand\
  O'Hagan}{2006}{Kennedy2006}
Kennedy, M.~C., Anderson, C.~W., Conti, S. \harvardand\ O'Hagan, A.
  \harvardyearleft 2006\harvardyearright , `Case studies in {G}aussian process
  modelling of computer codes', {\em Reliability Engineering \& System Safety}
  {\bf 91}(10-11),~1301--1309.

\harvarditem{Kennedy \harvardand\ O'Hagan}{2000}{Kennedy2000}
Kennedy, M. \harvardand\ O'Hagan, A.  \harvardyearleft 2000\harvardyearright ,
  `Predicting the output from a complex computer code when fast approximations
  are available', {\em Biometrika} {\bf 87}(1),~1--13.

\harvarditem{Kennedy \harvardand\ O'Hagan}{2001}{Ohagan01}
Kennedy, M. \harvardand\ O'Hagan, A.  \harvardyearleft 2001\harvardyearright ,
  `Bayesian calibration of computer models', {\em Journal Of The Royal
  Statistical Society Series B-Statistical Methodology} {\bf 63},~425--450.

\harvarditem[Kersting et~al.]{Kersting, Plagemann, Pfaff \harvardand\
  Burgard}{2007}{Kersting2007}
Kersting, K., Plagemann, C., Pfaff, P. \harvardand\ Burgard, W.
  \harvardyearleft 2007\harvardyearright , Most likely heteroscedastic
  {G}aussian process regression, {\em in} `Proceedings of the 24th
  international conference on Machine learning', ACM, pp.~393--400.

\harvarditem{Le~Gratiet \harvardand\ Cannamela}{2015}{Gratiet2015}
Le~Gratiet, L. \harvardand\ Cannamela, C.  \harvardyearleft
  2015\harvardyearright , `Cokriging-based sequential design strategies using
  fast cross-validation techniques for multi-fidelity computer codes', {\em
  Technometrics} {\bf 57}(3),~418--427.

\harvarditem{{Le Gratiet} \harvardand\ Garnier}{2014}{Le2014}
{Le Gratiet}, L. \harvardand\ Garnier, J.  \harvardyearleft
  2014\harvardyearright , `Recursive co-{K}riging model for design of computer
  experiments with multiple levels of fidelity', {\em International Journal for
  Uncertainty Quantification} {\bf 4}(5),~365--386.

\harvarditem[Marrel et~al.]{Marrel, Iooss, Da~Veiga \harvardand\
  Ribatet}{2012}{Marrel2012}
Marrel, A., Iooss, B., Da~Veiga, S. \harvardand\ Ribatet, M.  \harvardyearleft
  2012\harvardyearright , `Global sensitivity analysis of stochastic computer
  models with joint metamodels', {\em Statistics and Computing} {\bf
  22}(3),~833--847.

\harvarditem[{McKay} et~al.]{{McKay}, Beckman \harvardand\
  Conover}{1979}{Mckay1979}
{McKay}, M.~D., Beckman, R.~J. \harvardand\ Conover, W.~J.  \harvardyearleft
  1979\harvardyearright , `Comparison of three methods for selecting values of
  input variables in the analysis of output from a computer code', {\em
  Technometrics} {\bf 21}(2),~239--245.

\harvarditem{Oakley \harvardand\ O'Hagan}{2004}{Oakley04}
Oakley, J.~E. \harvardand\ O'Hagan, A.  \harvardyearleft 2004\harvardyearright
  , `{Probabilistic sensitivity analysis of complex models: a Bayesian
  approach}', {\em Journal of the Royal Statistical Society: Series B
  (Statistical Methodology)} {\bf 66}(3),~751--769.

\harvarditem{Oakley \harvardand\ {O'Hagan}}{2019}{Shelf4}
Oakley, J.~E. \harvardand\ {O'Hagan}, A.  \harvardyearleft
  2019\harvardyearright , `{SHELF: the Sheffield Elicitation Framework (version
  4)}'.
\newline\harvardurl{http://tonyohagan.co.uk/shelf}

\harvarditem{Oakley \harvardand\ Youngman}{2017}{Oakley2017}
Oakley, J.~E. \harvardand\ Youngman, B.~D.  \harvardyearleft
  2017\harvardyearright , `Calibration of stochastic computer simulators using
  likelihood emulation', {\em Technometrics} {\bf 59}(1),~80--92.

\harvarditem{O'Hagan}{2006}{Ohagan2006}
O'Hagan, A.  \harvardyearleft 2006\harvardyearright , `Bayesian analysis of
  computer code outputs: A tutorial', {\em Reliability Engineering \& System
  Safety} {\bf 91}(10-11),~1290--1300.

\harvarditem{O'Hagan}{2019}{Ohagan2019}
O'Hagan, A.  \harvardyearleft 2019\harvardyearright , `Expert knowledge
  elicitation: subjective but scientific', {\em The American Statistician} {\bf
  73}(sup1),~69--81.

\harvarditem{Overstall \harvardand\ Woods}{2016}{Overstall2016}
Overstall, A.~M. \harvardand\ Woods, D.~C.  \harvardyearleft
  2016\harvardyearright , `Multivariate emulation of computer simulators: model
  selection and diagnostics with application to a humanitarian relief model',
  {\em Journal of the Royal Statistical Society. Series C, Applied statistics}
  {\bf 65}(4),~483.

\harvarditem[Paterson et~al.]{Paterson, D'Amico, Thies, Kurt \harvardand\
  Harrison}{2018}{Paterson2018}
Paterson, J., D'Amico, F., Thies, P., Kurt, R. \harvardand\ Harrison, G.
  \harvardyearleft 2018\harvardyearright , `Offshore wind installation
  vessels--a comparative assessment for {UK} offshore rounds 1 and 2', {\em
  Ocean Engineering} {\bf 148},~637--649.

\harvarditem{Plumlee \harvardand\ Tuo}{2014}{Plumlee2014}
Plumlee, M. \harvardand\ Tuo, R.  \harvardyearleft 2014\harvardyearright ,
  `Building accurate emulators for stochastic simulations via quantile
  {K}riging', {\em Technometrics} {\bf 56}(4),~466--473.

\harvarditem{Rasmussen}{2006}{Rasmussen2006}
Rasmussen, C.~E.  \harvardyearleft 2006\harvardyearright , {\em Gaussian
  processes for Machine Learning}, Adaptive computation and machine learning,
  MIT Press, Cambridge, Mass.

\harvarditem[Rocchetta et~al.]{Rocchetta, Zio \harvardand\
  Patelli}{2018}{Rocchetta2018}
Rocchetta, R., Zio, E. \harvardand\ Patelli, E.  \harvardyearleft
  2018\harvardyearright , `A power-flow emulator approach for resilience
  assessment of repairable power grids subject to weather-induced failures and
  data deficiency', {\em Applied Energy} {\bf 210},~339--350.

\harvarditem{Rougier \harvardand\ Kern}{2010}{Rougier2010}
Rougier, J. \harvardand\ Kern, M.  \harvardyearleft 2010\harvardyearright ,
  `Predicting snow velocity in large chute flows under different environmental
  conditions', {\em Journal of the Royal Statistical Society: Series C (Applied
  Statistics)} {\bf 59}(5),~737--760.

\harvarditem[Sacks et~al.]{Sacks, Welch, Mitchell \harvardand\
  Wynn}{1989}{Sacks89}
Sacks, J., Welch, W.~J., Mitchell, T.~J. \harvardand\ Wynn, H.~P.
  \harvardyearleft 1989\harvardyearright , `Design and analysis of computer
  experiments', {\em Statistical Science} {\bf 4}(4),~409--423.

\harvarditem[Saisana et~al.]{Saisana, Saltelli \harvardand\
  Tarantola}{2005}{Saisana2005}
Saisana, M., Saltelli, A. \harvardand\ Tarantola, S.  \harvardyearleft
  2005\harvardyearright , `Uncertainty and sensitivity analysis techniques as
  tools for the quality assessment of composite indicators', {\em Journal of
  the Royal Statistical Society: Series A (Statistics in Society)} {\bf
  168}(2),~307--323.

\harvarditem[Santner et~al.]{Santner, Williams, Notz \harvardand\
  Williams}{2003}{Santner2003}
Santner, T.~J., Williams, B.~J., Notz, W. \harvardand\ Williams, B.~J.
  \harvardyearleft 2003\harvardyearright , {\em The {D}esign and {A}nalysis of
  {C}omputer {E}xperiments}, Vol.~1, Springer.

\harvarditem[Singh et~al.]{Singh, Couckuyt, Elsayed, Deschrijver \harvardand\
  Dhaene}{2017}{Singh2017}
Singh, P., Couckuyt, I., Elsayed, K., Deschrijver, D. \harvardand\ Dhaene, T.
  \harvardyearleft 2017\harvardyearright , `Multi-objective geometry
  optimization of a gas cyclone using triple-fidelity co-kriging surrogate
  models', {\em Journal of Optimization Theory and Applications} {\bf
  175}(1),~172--193.

\harvarditem{{Stan Development Team}}{2020}{stan}
{Stan Development Team}  \harvardyearleft 2020\harvardyearright , `{RStan}: the
  {R} interface to {Stan}'.
\newblock R package version 2.21.2.
\newline\harvardurl{http://mc-stan.org/}

\harvarditem{Sudret}{2008}{Sudret2008}
Sudret, B.  \harvardyearleft 2008\harvardyearright , `Global sensitivity
  analysis using polynomial chaos expansions', {\em Reliability Engineering \&
  System Safety} {\bf 93}(7),~964--979.

\harvarditem{Vanhellemont \harvardand\ Ruddick}{2014}{Vanhellemont2014}
Vanhellemont, Q. \harvardand\ Ruddick, K.  \harvardyearleft
  2014\harvardyearright , `Turbid wakes associated with offshore wind turbines
  observed with landsat 8', {\em Remote Sensing of Environment} {\bf
  145},~105--115.

\harvarditem[Wilson et~al.]{Wilson, Henderson \harvardand\
  Quigley}{2018}{Wilson2018}
Wilson, K.~J., Henderson, D.~A. \harvardand\ Quigley, J.  \harvardyearleft
  2018\harvardyearright , `Emulation of utility functions over a set of
  permutations: sequencing reliability growth tasks', {\em Technometrics} {\bf
  60}(3),~273--285.

\harvarditem[Zitrou et~al.]{Zitrou, Bedford \harvardand\ Walls}{2016}{Zit16}
Zitrou, A., Bedford, T. \harvardand\ Walls, L.  \harvardyearleft
  2016\harvardyearright , `A model for availability growth with application to
  new generation offshore wind farms', {\em Reliability Engineering and System
  Safety} {\bf 152}(C),~83--94.

\harvarditem[Zitrou et~al.]{Zitrou, Bedford, Walls, Wilson \harvardand\
  Bell}{2013}{Zit13}
Zitrou, A., Bedford, T., Walls, L., Wilson, K. \harvardand\ Bell, K.
  \harvardyearleft 2013\harvardyearright , Availability growth and
  state-of-knowledge uncertainty simulation for offshore wind farms, {\em in}
  `22nd ESREL conference 2013'.
\newline\harvardurl{https://strathprints.strath.ac.uk/45377/}

\end{thebibliography}
\bibliographystyle{agsm}	
\end{document}